 \documentclass[preprint2]{aastex}

\newcommand{\kms}{km~s$^{-1}$}

 \slugcomment{Draft version 2012} 

\shorttitle{A 3D view of GK~Per}
\shortauthors{Liimets et al.}

\begin{document}


\title{A three-dimensional view of the remnant of Nova Persei 1901 (GK~Per)}


\author{T. Liimets\altaffilmark{1,2},
  R.L.M. Corradi\altaffilmark{3,4},
  M. Santander--Garc{\'{\i}}a\altaffilmark{5,6},
  E. Villaver\altaffilmark{7}, 
  P. Rodr{\'{\i}}guez-Gil\altaffilmark{3,4},
  K. Verro\altaffilmark{1,2}, I. Kolka\altaffilmark{1}}


\altaffiltext{1}{Tartu Observatory, T\~oravere, 61602, Estonia}
\altaffiltext{2}{Institute of Physics, University of Tartu, Riia 142, 51014, Estonia}
\altaffiltext{3}{Instituto de Astrof{\'{\i}}sica de Canarias, V\'\i a L\'actea s/n, La Laguna, E-38205, Santa Cruz de Tenerife, Spain}
\altaffiltext{4}{Departamento de Astrof{\'{\i}}sica, Facultad de F\'\i sica y Matem\'aticas, Universidad de La Laguna, Avda. Astrof\'\i sico Francisco S\'anchez s/n, La Laguna, E-38206, Santa Cruz de Tenerife, Spain}
\altaffiltext{5}{Observatorio Astron\'omico Nacional, Ap.\ de Correos 112, E-28803, Alcal\'a de Henares, Madrid, Spain}
\altaffiltext{6}{Centro de Astrobiolog\'\i a, CSIC-INTA, Ctra de Torrej\'on a Ajalvir km 4, E-28850 Torrej\'on de Ardoz, Spain}
\altaffiltext{7}{Departamento de F\'{\i}sica Te\'orica, Universidad
Aut\'onoma de Madrid, E-28049 Madrid, Spain}


\begin{abstract}

We present a kinematical study of the optical ejecta of GK~Per.  It
is based on proper motions measurements of 282 knots from
$\sim$20 images spanning 25 years. Doppler-shifts are also computed for 217 knots.  The
combination of proper motions and radial velocities allows a unique
3-D view of the ejecta to be obtained. The main results are:
(1) the outflow is a thick shell in which knots expand with a significant range of
velocities, mostly between 600 and 1000~\kms;
(2) kinematical ages indicate that knots have suffered only a
modest deceleration since their ejection a century ago;
(3) no evidence for anisotropy in the expansion rate is found;
(4) velocity vectors are generally aligned along the radial direction
but a symmetric pattern of non-radial velocities is also observed at
specific directions;
(5) the total $H\alpha$+[N~II] flux has been linearly
decreasing at a rate of 2.6\%\ per year in the last decade. 
The Eastern nebular side is fading at a slower rate than 
the Western one. Some of the knots displayed a rapid change of brightness during the 2004-2011
period. Over a longer timescale, a progressive circularization and homogenization of the nebula
is taking place;
(6) a kinematic distance of 400$\pm$30~pc is determined.
These results raise some problems to the previous interpretations of
the evolution of GK~Per. In particular, the idea of a strong
interaction of the outflow with the surrounding medium in the
Southwest quadrant is not supported by our data. 

\end{abstract}



\section{INTRODUCTION}

Nova Persei 1901 (GK~Per) was discovered by Anderson in 1901 February 21  
at a visual magnitude of 2.7.  Only one day before, its brightness
must have been fainter than 12 mag \citep{w01}. It reached the maximum
light on February 22 when it was as bright as Vega (0.2 mag). GK~Per
was the first object around which superluminal light echoes were
detected (e.g. \citealt{r02}). The actual ejecta of the nova outburst
was discovered in 1916 by \citet{bar16} and is still easily observable
now: indeed, it is considered to be the longest lived and most
energetic nova remnant ever found. 

The earliest optical images of the remnant showed an emission only in
the Southwest (SW) side of the remnant
(e.g. \citealt{seaq89}). Gradually more material became visible around
the central source, however, the asymmetric nature of the remnant has
remained.
Radio observations obtained during 1984--1986 showed that the SW
quadrant of the remnant is the source of non-thermal (synchrotron)
emission \citep{seaq89}. This would be caused by particle acceleration
and enhanced magnetic fields in a fast shock where the expelled matter
is meeting some pre-existing material, most probably the result of
previous mass ejections from the system, possibly an ancient planetary
nebula (\citealt{bode04}).
This scenario is supported by the analysis of the X-ray emission from
the remnant \citep{bal05}.  A lumpy, highly asymmetric X-ray nebula is
observed, with brightness enhancement in the SW quadrant and wing-like
extensions in the Southeast (SE) and Northwest (NW) directions. It is
concluded that a density gradient in the circumstellar medium causes
the expansion of the X-ray shell to be faster in the NW and SE
directions, reaching velocities as high as $\sim$2600 - 2800~\kms,
than toward the SW, where velocities are less than half of that,
around $\sim$1100~\kms.  This strong interaction is expected to cause
deceleration of the expansion of the SW part of the outflow
\citep{bode04}.


At a distance of half a kiloparsec (e.g. \citealt{mcl60}) and with an
expansion velocity of a thousand \kms, the apparent growth of the
nebula is about one arcsec per year, which is easily resolvable from
ground-based optical imagery even on a timescale of few
months. Therefore GK~Per offers the opportunity to study the evolution
of a stellar outflow in real-time, and to determine not only the
expansion velocities but also their derivatives (i.e. acceleration,
thus the forces at work), or deviations from pure radial expansion, an
important information that is generally not available for dynamical
studies of astrophysical outflows. If the expansion in the plane of
the sky is coupled with Doppler shift velocities, information about
five of the six dimensions of the phase-space is gained, and an almost
complete dynamical picture can be drawn.
For these reasons, starting on 2004 we embarked in a program of
frequent imaging monitoring of the expansion of the GK~Per remnant. We
present here the results of the work.  Doppler shift velocities of 217
knots
from long-slit spectroscopy are combined with high-precision
measurements of the motion in the plane of the sky for 282 knots. The
latter ones are computed using good-quality, ground-based $H\alpha$+[NII]
images spanning 25 years, with a particularly intensive time coverage
since 2004.

At the beginning of the article we describe our data and archival
data.  Then designations used in this article are given
(Section~\ref{S-geom}).  Proper motions and radial velocities for
individual knots are discussed in Section~\ref{S-pm} and \ref{S-rv},
and the kinematical analysis is presented in Section~\ref{S-kinmod}.
Section~\ref{S-br} presents the brightness evolution of the nebula,
and Section~\ref{S-disc} presents the discussion.

\section{OBSERVATIONS AND DATA REDUCTION}

\subsection{New Imaging}

Most of the imaging was carried out at the 2.5m Isaac Newton Telescope
(INT) at the Observatorio del Roque de los Muchachos (ORM), La Palma,
Spain. Images were taken with the Wide Field Camera (WFC) and a narrow
band $H\alpha$ filter having a central wavelength of 6568~\AA\ and a
bandpass of 95~\AA.  The pixel scale is $0''.33$ pix$^{-1}$.

Other images were taken with the Andalucia Faint Object
Spectrograph and Camera (ALFOSC) at the 2.6m Nordic Optical Telescope
(NOT) of the ORM. An $H\alpha$ filter (6577/180~\AA) was used, and the
pixel scale was $0''.189$ pix$^{-1}$.
Both the INT and NOT filters are broad enough to include the emission
from the [NII] 6548~\AA\ and 6583~\AA\ doublet, which are the
strongest optical emission lines in the ejecta, and to cover its whole
radial velocity range ($\pm1000$~\kms).

All images were obtained with a seeing better than 1.3 arcsec
full-width at half-maximum (FWHM).

\subsubsection{Archival imaging}

We also searched for archival images of the nebula. INT images
obtained on 1999 \citep{bode04} with the WFC, and on 1987
\citep{seaq89} with the Prime Focus Cone Unit (PFCU, pixel scale
$0''.74$ pix$^{-1}$) were downloaded from the Isaac Newton Group
archive. The 1999 data have been obtained with the same filter as our
recent INT+WFC observations. The 1987 frame was observed in a
broadband R filter.

The [NII] Hubble Space Telescope (HST) 1995 and 1997 images discussed
by \citet{sh12}
were downloaded from the MAST public archive. 
They were obtained with the Wide Field Planetary Camera 2 (WFPC2): the
relatively large remnant fills the field of view of the camera, and is
sampled at a $0''.1$ pix$^{-1}$ by the WF detectors, and $0''.045$
pix$^{-1}$ by the PC CCD.  In both epochs, the filter F658N was used,
which has a central wavelength 6590~\AA\ and a bandpass of
22~\AA. This is narrower than the range covered by the ejecta and some
knots with the most extreme radial velocity shifts are missed.

The Digitized Sky Survey (DSS) images from the Palomar 48-inch Samuel
Oschin telescope (Pal48) obtained on 1953 (pixel scale $1''.7$
pix$^{-1}$) and 1989 (pixel scale $1''.0$ pix$^{-1}$) were retrieved
via the Aladin interface.  In both cases observations were done on
red-sensitive photographic plates with a bandpass approximately
equivalent to an R filter.  In 1953 a plate Kodak 103a-E and in 1989 a
IIIaF was used.  The 1953 image is used for visual comparison with the
recent images, but not used for any quantitative measurements. The
1989 image could instead be used to measure the position of several
knots recognized in our most recent imagery.

The log of all the imaging observations is presented in
Table~\ref{T-imaobs}.
\begin{deluxetable}{llllrcc}
\tabletypesize{\scriptsize}
\tablecaption{Log of the imaging data\label{T-imaobs} }
\tablewidth{0pt}
\tablehead{
\colhead{Date} & \colhead{JD} & \colhead{Telescope+}  & \colhead{Filter}                      & \colhead{Total exp.}   & \colhead{Number}    & \colhead{Original seeing}    \\ 
\colhead{}         &  \colhead{}     &  \colhead{Instrument}    & \colhead{CW/FWHM (\AA)} &  \colhead {time (sec)}  &  \colhead {of frames}  &  \colhead {(arcsec)}   
}
\startdata
1953-01-10        & 2434387.62 & Pal48 & phot. plate        & 2400 & 1 & 5.0       \\
1987-07-27\tablenotemark{a}     & 2447003.69 & INT+PFCU & R & 300  & 1 & 2.8                    \\
1989-10-05        & 2447804.87 & Pal48 &phot. plate         & 5100 & 1 & 4.9             \\
1995-11-08\tablenotemark{b} & 2450030.42 & HST+WFPC2 & 6590/22 & 1400 & 2 &0.2             \\
1997-01-08\tablenotemark{b} & 2450457.18 & HST+WFPC2 & 6590/22 & 1200 & 2 & 0.2             \\
1999-11-30$^\ast$\tablenotemark{c} & 2451513.56 & INT+WFC   & 6568/95   &  7026 & 3 & 2.3          \\
\bf{2004-01-29}$^\ast$ & 2453034.37 & INT+WFC & 6568/95     & 900  & 3  & 1.2           \\
2004-08-04$^\ast$ & 2453222.66 & INT+WFC  & 6568/95         & 1500 & 5  & 1.2        \\
2004-09-03$^\ast$ & 2453252.67 & INT+WFC  & 6568/95         & 1500 & 5  & 1.1        \\
2004-12-30$^\ast$ & 2453370.43 & INT+WFC  & 6568/95         & 1200 & 4  & 1.1         \\
2005-09-16$^\ast$ & 2453630.72 & INT+WFC  & 6568/95         & 900  & 3  & 1.0        \\
2006-10-12$^\ast$ & 2454021.53 & INT+WFC  & 6568/95         & 1200 & 4  & 1.0        \\
2007-01-04$^\ast$ & 2454105.49 & INT+WFC  & 6568/95         & 300  & 1  & 1.0        \\
2007-03-01$^\ast$ & 2454161.42 & INT+WFC  & 6568/95         & 2100 & 7  & 1.2        \\
2007-09-03        & 2454347.68 & NOT+ALFOSC & 6577/180      & 1800 & 3  & 0.6        \\
2007-09-05        & 2454349.71 & NOT+ALFOSC & 6577/180      & 1800 & 3  & 0.5        \\
2007-09-07$^\ast$ & 2454351.74 & INT+WFC  & 6568/95         & 1200 & 4  & 1.1        \\
2008-01-08$^\ast$ & 2454474.42 & INT+WFC  & 6568/95         & 1200 & 4  & 0.8        \\
2008-08-19$^\ast$ & 2454698.69 & INT+WFC  & 6568/95         & 4650 & 31 & 1.1        \\
2008-09-12        & 2454722.70 & INT+WFC  & 6568/95         & 1200 & 4  & 1.3               \\
2008-11-14$^\ast$ & 2454785.49 & INT+WFC  & 6568/95         & 1500 & 5  & 1.2        \\
2009-02-11        & 2454874.43 & NOT+ALFOSC  & 6577/180     & 1200 & 2  & 0.8       \\
2009-12-08        & 2455174.48 & INT+WFC  & 6568/95         & 300  & 1  & 1.3            \\
2010-02-09$^\ast$ & 2455237.37 & INT+WFC  & 6568/95         & 1500 & 5  & 0.9        \\
2011-12-13$^\ast$ & 2455909.41 & INT+WFC  & 6568/95         & 2700 & 9  & 0.8        \\
\enddata
\tablecomments{The date indicates the start of observing night in
  the format yyyy-mm-dd. The date marked in boldface corresponds to the image 
  that was used as reference for the astrometry and flux match of all images.
Images from all dates marked with $^\ast$ were flux and PSF matched with the 
reference image. For 1999 only flux match was done due to its modest seeing. 
JD is the Julian Date at mid point of observations. In column 6 is the number of frames added together.}
\tablenotetext{a}{Published in \citet{seaq89}}
\tablenotetext{b}{Published in \citet{sh12}}
\tablenotetext{c}{Published in \citet{bode04}}
\end{deluxetable}

\subsubsection{Imaging data reduction}\label{S-imdr}

The basic reduction (bias, flat field, combination of sub-exposures,
and cosmics removal) of most of the images was done using standard routines
in IRAF\footnote{IRAF is distributed by the National Optical Astronomy 
Observatory, which is operated by the Association of Universities
for Research in Astronomy (AURA) under cooperative agreement with
the National Science Foundation.}. 
HST data were reduced using tasks in the package {\tt dither}. 
All the frames were cropped to the same field of view (FOV) of 6.4x6.4
arcmin$^2$ around the central star, except for the data from 1987, 1995, and 1997 which 
originally had a smaller FOV.

The next step was to carefully register all images to the same
astrometric reference frame using stars from USNO-B1.0 catalogue with
proper motion errors smaller than $0''.002$ yr$^{-1}$. About sixty stars
were used in every frame, except in the older frames which had worse
seeing and smaller FOV. Our first INT image, obtained on 20040129, was
chosen to be the reference frame.
The Python based Kapteyn Package was used to get the reference frame
FITS header world coordinate system applied pixel-by-pixel to the
reference frame.  The other images were pixel-by-pixel matched to the
reference frame using tasks {\it wcsxymatch}, {\it geomap}, 
and {\it geotran} in IRAF.  The latter put all the frames in the same pixel
scale of the reference image, $0''.33$ pix$^{-1}$. Results were
checked using some twenty stars in the proximity or inside the
remnant. The standard deviation of the position of these stars from
the reference frame in the $X$ and $Y$ directions is 0.08 arcsec except
for the images taken before 2004 where it is occasionally slightly
larger. We adopt this figure as representative of the uncertainty
introduced by the astrometric registration of images of different
epochs.

To investigate brightness variations of the remnant, 
all the INT+WFC frames
marked with $^\ast$ in Table~\ref{T-imaobs}, were also flux
matched. First, all the frames were degraded down to seeing of the
reference frame, 1.2 arcsec. Secondly, frames were flux matched using
forty-eight well exposed stars in the FOV. No stars inside the remnant
were used to avoid possible contamination from the moving knots.  
The rms of fluxes of the reference stars on different frames is 1.3~\%.

\subsection{Spectroscopy}

Medium-resolution, long-slit spectra of the nebula were obtained
during two observing periods.  In 2007 January, the INT+IDS
(Intermediate Dispersion Spectrograph) with grating R1200Y and a
$1''.5$ slit was used. This configuration covers a spectral range from
5730~\AA\ to 7610~\AA\ at a dispersion of 0.47~\AA~pix$^{-1}$.  In
2007 September, the NOT+ALFOSC with grism~\#17 and $0''.5$ slit was
used, giving spectral range from 6350~\AA\ to 6850~\AA, and a
dispersion of 0.26~ \AA~pix$^{-1}$. 
The spectrographs slit was oriented at a dozen of different position
angles with the aim to cover as many knots as possible. Details of the
spectroscopic observations, including the orientation of the slit, 
the exposure times, and the spectral resolution 
with the adopted slit widths, can be found in Table~\ref{T-specobs}. 
These data were reduced and wavelength calibrated using the standard
routines in the package {\tt longslit} in IRAF.

\begin{deluxetable}{lcccccc}
\tabletypesize{\scriptsize}
\tablecaption{Log of the spectroscopic observations\label{T-specobs}}
\tablewidth{0pt}
\tablehead{
\colhead{Date} & \colhead{Telescope+} & \colhead{Slit P.A.} & \colhead{Total  exp.}  & \colhead{Number} &  \colhead{Resolution} &  \colhead{Spectral range} \\
\colhead{}     & \colhead{Instrument} & \colhead{($\degr$)}                  &  \colhead{time (sec)}  &  \colhead {of frames} & \colhead{(\AA)}  & \colhead{(\AA)}          
}
\startdata

2007-01-13 & INT+IDS         &  22.5 & 3600 & 2 & 1.3 & 5730 -- 7610\\
                   &                             &  48   & 3600  & 2  &     & \\
                   &                             &  315  & 1800 & 1 &     & \\
2007-01-14 & INT+IDS                             &  69   & 1800 & 1 & 1.3  &5730 -- 7610\\
                   &                             &  293  & 3600 & 2 &    & \\
                   &                             &  334  & 3600 & 2 &     & \\
2007-09-03 & NOT+ALFOSC                          &  31   & 3600 & 2 & 0.7  & 6350 -- 6850\\
2007-09-04 & NOT+ALFOSC                          &  86   & 1800 & 1 & 0.7 & 6350 -- 6850 \\
                   &                             & 107   & 1500 & 1 &    &  \\
                   &                             &  312  & 1800 & 1 &     & \\
2007-09-05 & NOT+ALFOSC                          &  49   & 1800 & 1 & 0.7 & 6350 -- 6850 \\
                   &                             &  173  & 1800 & 1 &   &  \\
\enddata
\tablecomments{P.A. is measured from North to East. 
Column 5 indicates a number of frames added together. See text for more details.}
\end{deluxetable}

\section{DESCRIPTION OF THE NEBULA}\label{S-dn}

\begin{figure*}[!t]
\epsscale{2}
\plotone{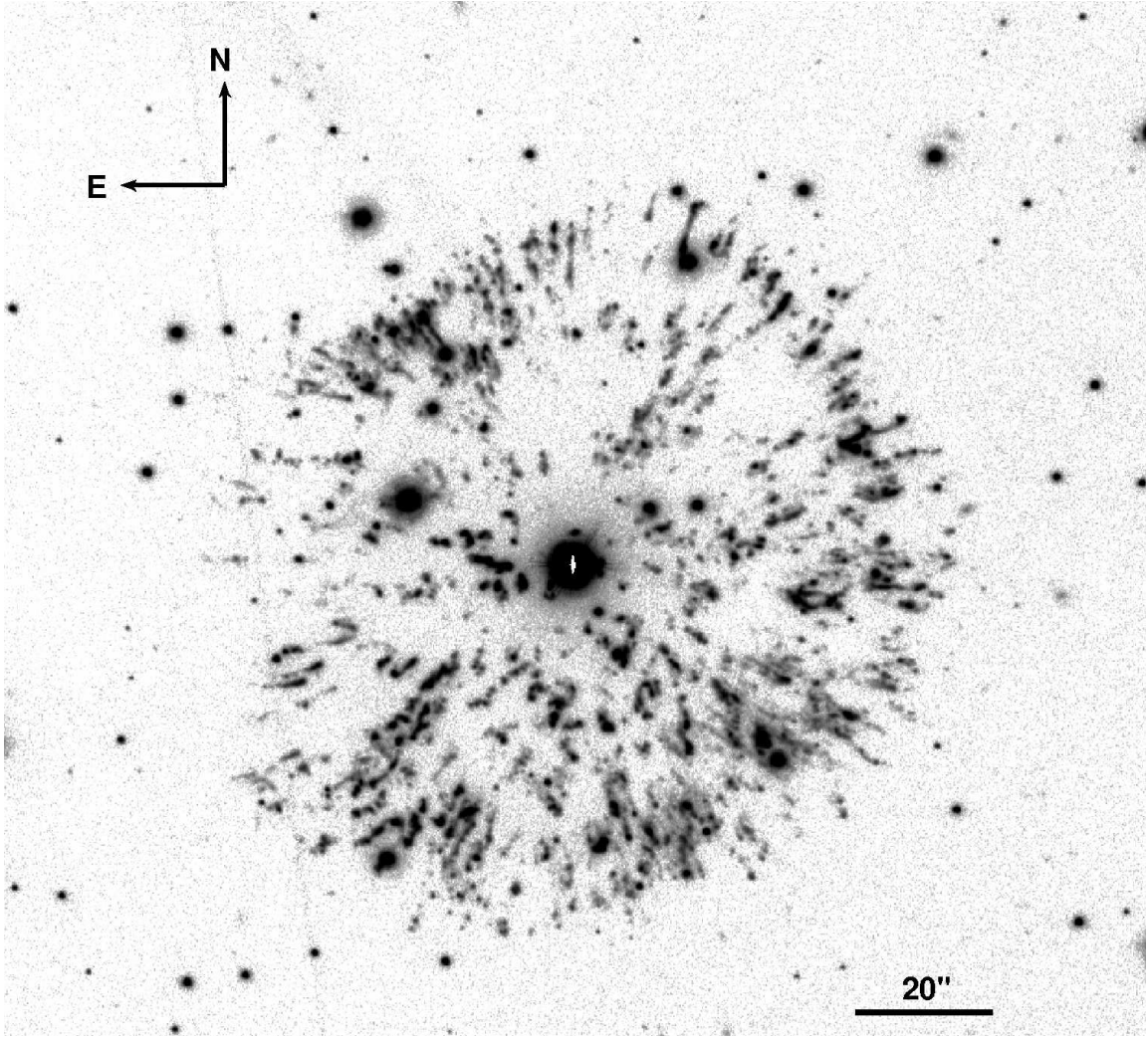}
\caption{Original frame from 20070905 obtained with the NOT. The image
  shows the filamentary nature of the GK~Per remnant. The FOV is
  2.8x2.8 arcmin$^2$. \label{F-image}}
\end{figure*}

The optical nebula of GK~Per, displayed in Figure~\ref{F-image}, has a
roundish, knotty morphology. Some deviation from the circular symmetry
is however visible because the diameter of the nebula in the NE-SW
direction is smaller ($105''$ as measured in the 2011 image) than in
the NW-SE direction ($118''$).  As described in details by
\citet{sh12}, the nebula is composed of hundreds of knots and
filaments of different sizes and brightnesses. Many of the knots have
tails pointing either toward or away from the central star. 
\citet{sh12} counted 937 knots in their HST images. Some of them are
blended in our lower-resolution ground-based images. On the other
hand, the limited bandpass of the [NII] F658N HST filter makes that
some knots with the largest Doppler-shift displacements, which
owing to the projection effects are located near the center of the
remnant, are missed in the HST images.

\section{GEOMETRICAL DESIGNATIONS}\label{S-geom}

\begin{figure}[!t]
\epsscale{1}
\plotone{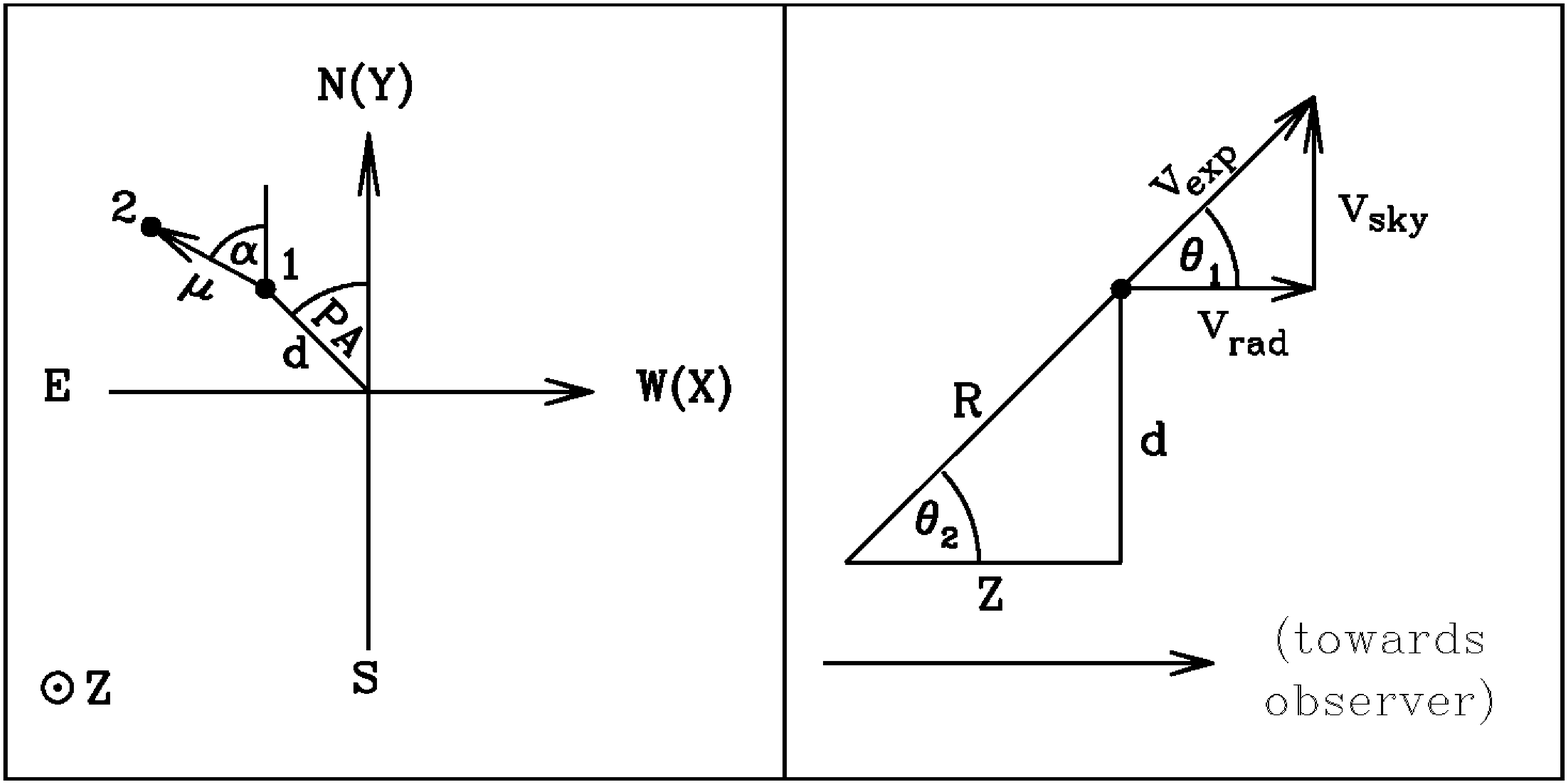}
  \caption{Designations used through this article. Left panel: sketch
    of a example knot as seen in the plane of the sky. Right panel:
    view in the perpendicular plane including the central star and the
    knot. See text for more details. \label{F-geom}}
\end{figure}

Throughout this article we use the designations as illustrated in
Figure~\ref{F-geom}.  In its left panel, we plot a view in the plane
of the sky, where the line of sight corresponds to the $Z$ axis and is
perpendicular to the plane of the paper. The $X$ axis points toward
West, and the $Y$ axis to North (up).  Points 1 and 2 indicate a
knot's position at two different epochs. The displacement over one
year (proper motion) is indicated by $\mu$ [arcsec yr$^{-1}$], and its
angle in the plane of the sky is $\alpha$.  PA indicates the position
angle of the knot with respect to the central star as measured on our
20040129 frame. Both $\alpha$ and PA are measured from North to East
counterclockwise. The apparent distance of the knot from the central
star in the plane of the sky is named as $d$.

When deprojecting the observed positions, each knot has a spatial
distance $R$ from the central source (right panel of the
Figure~\ref{F-geom}). The plane indicated in the figure is defined by
the central star, the observer, and the knot considered: it is
therefore perpendicular to the XY plane, but changes depending on
the knot under consideration.
The speed of the knot is indicated by $v_{exp}$, and its components
along the line of sight (radial velocity) and in the plane of the sky
(tangential velocity) are indicated by $v_{rad}$ and $v_{sky}$,
respectively.  The angle between the line of sight and the velocity
vector is $\theta_{1}$.  The angle that the line joining the central
star with the knot forms with the line of sight is $\theta_{2}$: in
the case of purely radial expansion, $\theta_{1}=\theta_{2}$. Both
angles are measured from the line of sight toward the velocity vector. 

\begin{figure}[!b]
\plotone{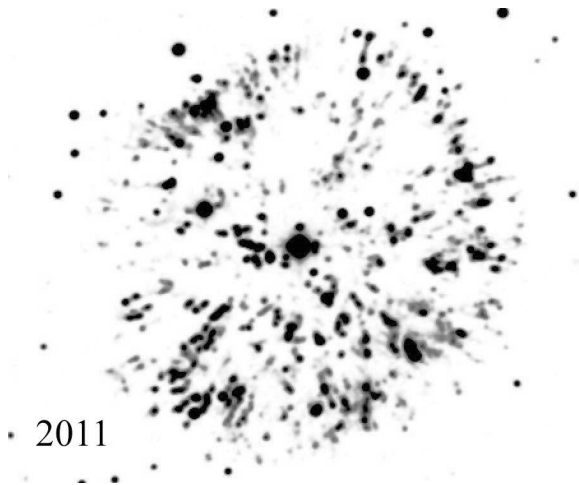}
\caption{Animation of the evolution of the GK~Per optical nebula from
    1953 to 2011. This animation is only available in the electronic
    version of the article. For arXiv version see ancillary file figure3\_animation.mov .
\label{F-video}}
\end{figure}

\section{PROPER MOTIONS}\label{S-pm}

Our multi-epoch imaging allows a precise determination of the apparent
expansion of the GK~Per nebula in the plane of the sky. This is
illustrated in the video in Figure~\ref{F-video} (only available in
the electronic version of the article). The animation includes the
early photographic and CCD images, astrometrically matched to the
recent ones but in an arbitrary intensity scale and with a generally
poor resolution, as well as a selection of the best INT images in the
sequence of highly homogeneous data taken starting on 2004, in the
same linear intensity scale. In the latter part of the sequence, not
only the overall expansion of the outflow is clearly revealed, but
also changes in brightness of some of the knots, as it will be
discussed in the Section~\ref{S-br}.

\begin{figure}[!h]
\plotone{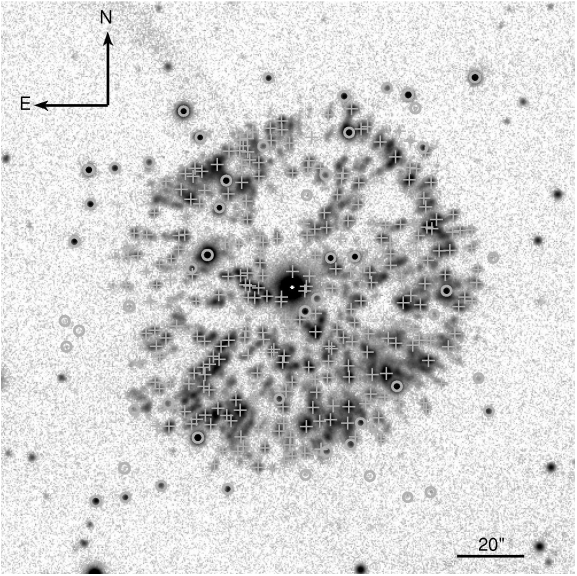}
 \caption{Knots with proper motion measurements (crosses, blue in online version) 
  indicated on our reference frame 20040129.  Circles (blue in online version) 
  show stars inside or nearby the remnant. FOV 2.8x2.8 arcmin$^2$. \label{F-2004pm} }
\end{figure}

\begin{figure}[!t]
\plotone{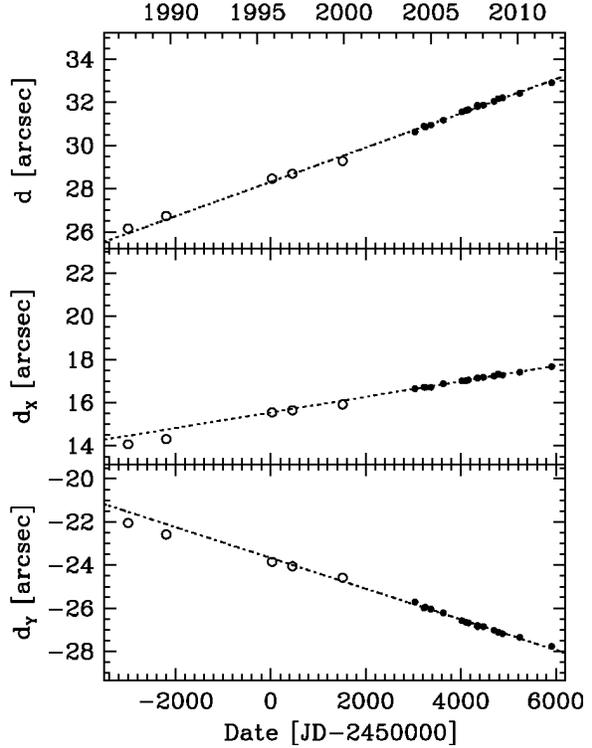}
  \caption{Example of proper motion determination of a knot.  The 
  upper panel shows the variation of distance $d$ with time, 
and the lower two panels its components along the
 $X$ and $Y$ axes ($d_{x}$ and $d_{y}$) as defined in Figure~\ref{F-geom}.
Open circles are data from 1987 to 1999 (not used for the proper
motion determination), and filled circles those obtained from 2004 to
2011.  Dotted lines are the least square fits of data
taken after 2004. Error bars are smaller or equal to sizes of symbols.\label{F-pm}}
\end{figure}
\begin{figure*}[!hp]
\epsscale{2.1}
\plotone{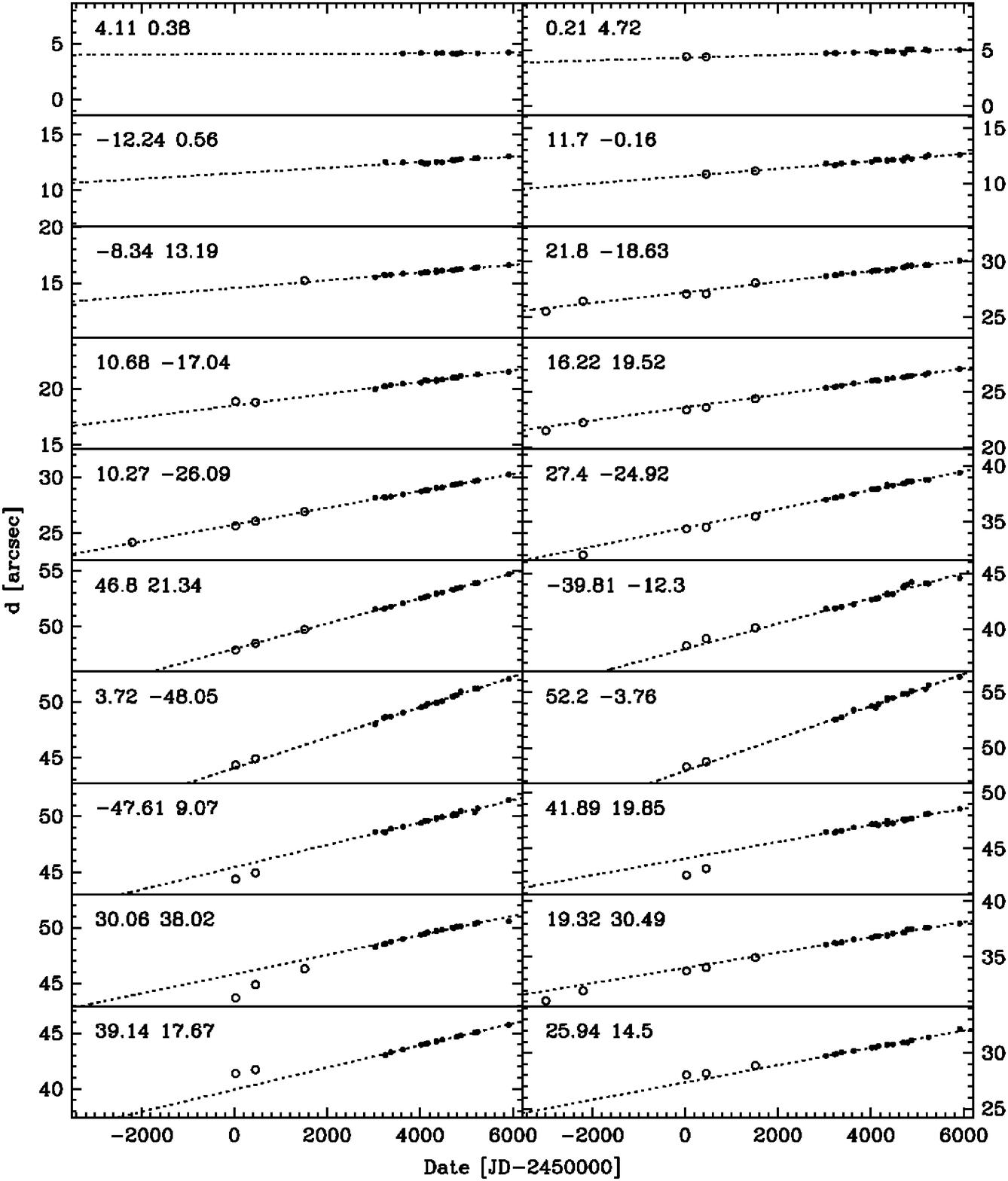}
\caption{More examples of proper motion determinations. 
In the first seven rows, knots are ordered according to increasing
proper motion magnitude (the range spanned by abscissae is the same
for all graphs, 10 arcsec). In the three bottom rows, some cases with
signs of deceleration/acceleration in the last two decades are
shown. At the upper-left corner of each panel, the knot's coordinates
($d_{x}$ and $d_{y}$ distances from the center, in arcsec, extrapolated at
the reference epoch, see Table~\ref{T-propm}) are indicated. Error bars are smaller or 
equal to sizes of symbols. See the caption of Figure~\ref{F-pm} for the explanation 
of legends. \label{F-pm_more}}
\end{figure*}

The motions in the plane of the sky for 282 individual knots were
determined using the nineteen, carefully registered images obtained
between 2004 and 2011. The measured knots are indicated in
Figure~\ref{F-2004pm}. The position of knots at each epoch was
computed by fitting independently in the $X$ and $Y$ directions a Gaussian
using the IRAF task {\it imexamine}, with fitting parameters
adjusted to the mean characteristic of the knots. Only knots for which
a reasonable Gaussian fit can be obtained were considered.
Then the proper motion of each knot were determined by means of a
least square fit of the positions at the different epochs. Both the
distance variation from the central star, as well the separate motions
in the $X$ and $Y$ directions were computed, as illustrated in
Figure~\ref{F-pm}.  In this way, the proper motion vector is determined,
which in addition to the magnitude of the spatial displacement $\mu$
also contains the information on its direction (angle
$\alpha=\arctan(\mu_{x}/\mu_{y})$, see Figure~\ref{F-geom}).
For all of the knots, a straight line provides an excellent fit to the variation of
$d$ during the 7.9-yr lapse of time considered. This is further
illustrated in Figure~\ref{F-pm_more}.  Proper
motions range from $0".007$ yr$^{-1}$ to $0."53$ yr$^{-1}$.

The prime focus camera of the INT has well defined and small geometrical
distortions\footnote{http://www.ast.cam.ac.uk/$\sim$wfcsur/technical/astrometry/}
in the limited FOV covered by the GK~Per nova remnant, and which are
robustly removed during the astrometric registering of
images. Therefore, errors in the proper motion determination mainly
depend on the goodness of the fit of the position of the knots at the
different epochs, which in turn depends on the knot's shape and
brightness. As the proper motion error, we adopt the formal error of
the least square fit: its average value is $0".010\pm0".006$ yr$^{-1}$.

In the proper motion calculation, we did not use the images earlier
than 2004 as they come from a less homogeneous set of observations
with generally poorer resolution or astrometric properties, as well as
different filters. They are however used to search for signs of
acceleration/deceleration of knots in the last decade with respect to
the previous 20 years. In this respect, we found no evidence for a
systematic acceleration/deceleration, except in a few cases (see
the lower three panels in Figure~\ref{F-pm_more}). Among these, a
modest slow down is the most common case.

\begin{deluxetable}{ccccccccccccc}
\tablecaption{Knots positions, proper motions, and radial velocities\label{T-propm}}
\tabletypesize{\tiny}
\tablewidth{0pt}
\tablehead{
\colhead{$d_x\tablenotemark{a}$} & \colhead{$d_y\tablenotemark{a}$} & \colhead{$\mu$}        & \colhead{$\sigma_{\mu}$} & \colhead{$\mu_{x}$}     & \colhead{$\sigma_{\mu_{x}}$} & \colhead{$\mu_{y}$}     & \colhead{$\sigma_{\mu_{y}}$} & \colhead{PA}          & \colhead{$\sigma_{PA}$} & \colhead{$\alpha$} & \colhead{$\sigma_{\alpha}$} & \colhead{$v_{rad}$} \\
\colhead{($"$)}        & \colhead{($"$)}        & \colhead{($"~yr^{-1}$)} & \colhead{($"~yr^{-1}$)}  & \colhead{($"~yr^{-1}$)} & \colhead{($"~yr^{-1}$)}     & \colhead{($"~yr^{-1}$)} & \colhead{($"~yr^{-1}$)}     &   \colhead{($\degr$)} & \colhead{($\degr$)}    &   \colhead{($\degr$)} &   \colhead{($\degr$)} &   \colhead{(km~s$^{-1}$)}
}
\startdata 
  0.06 & -27.43 &  0.242 & 0.005 &  0.005 & 0.004 & -0.242 & 0.005 &  180.1 &  0.3 &  181.2 &    0.9 & -- \\  
  0.09 & -48.91 &  0.448 & 0.006 & -0.009 & 0.012 & -0.448 & 0.006 &  180.1 &  0.2 &  178.8 &    1.5 & -- \\  
  0.21 &   4.72 &  0.046 & 0.011 & -0.007 & 0.012 &  0.046 & 0.011 &  357.5 &  2.0 &    8.9 &   14.7 &    808 \\  
  0.21 &   4.72 &  0.046 & 0.011 & -0.007 & 0.012 &  0.046 & 0.011 &  357.5 &  2.0 &    8.9 &   14.7 &    919 \\  
  0.22 &  33.44 &  0.261 & 0.003 & -0.005 & 0.004 &  0.261 & 0.003 &  359.6 &  0.3 &    1.0 &    0.8 & -- \\  
  2.25 &  -9.25 &  0.081 & 0.004 &  0.010 & 0.010 & -0.080 & 0.005 &  193.7 &  1.0 &  186.8 &    7.1 &   -822 \\  
  3.41 &  18.67 &  0.153 & 0.013 &  0.003 & 0.019 &  0.155 & 0.014 &  349.7 &  0.5 &  358.9 &    7.1 &    721 \\  
  3.69 &  -0.88 &  0.048 & 0.017 &  0.041 & 0.015 & -0.033 & 0.022 &  256.6 &  2.5 &  231.4 &   21.0 &   -894 \\  
  3.72 & -48.07 &  0.497 & 0.012 &  0.057 & 0.007 & -0.494 & 0.012 &  184.4 &  0.2 &  186.5 &    0.9 & -- \\  
  4.11 &   0.38 &  0.007 & 0.006 &  0.006 & 0.006 &  0.007 & 0.011 &  275.3 &  2.3 &  317.3 &   54.0 &   -794  \\ 
\enddata
\tablecomments{The entire table is published only in the electronic
  edition of the article (For arXiv version see ancillary file tabel3\_online.txt).  The first 10 lines are shown here for
  guidance regarding its form and content.}
\tablenotetext{a}{Extrapolated to 2004-01-29, or JD2453034.37.}
\end{deluxetable} 

The knots positions, their proper motions $\mu$, $\mu_{x}$, $\mu_{y}$,
and direction $\alpha$, position angles, radial velocities, and the
errors associated to all these quantities, are listed in
Table~\ref{T-propm}. An excerpt is presented in the paper, while the
entire table is only available in the electronic version of the
article.  In the table, individual knots are identified by their X and
Y distance in arcsec from the central source ($d_{x}$, $d_{y}$) on the
plane of the sky at the reference epoch as determined from the adopted
least square fits.

\subsection{Comparison with published data}

A study of the proper motions in GK Persei remnant knots was recently
done by \citet{sh12} using the 1995 and 1997 HST images listed in
Table~\ref{T-imaobs}.  The proper motions of 938 knots is estimated,
but no quantitative comparison with our measurements is possible as
\citet{sh12} did not publish data for individual knots.  In spite of
the higher resolution of the HST, the seven-times longer baseline of our
observations, the large number of measurements (19, which reduce
statistical errors), and the accurate knowledge of the astrometric
properties of the INT prime-focus camera, provide proper motion
measurements which are more accurate by a factor of four (cf. Figure~8
in \citealt{sh12} with our average error of 0."01~yr$^{-1}$).

\begin{figure}[!h]
\epsscale{1.0}
\plotone{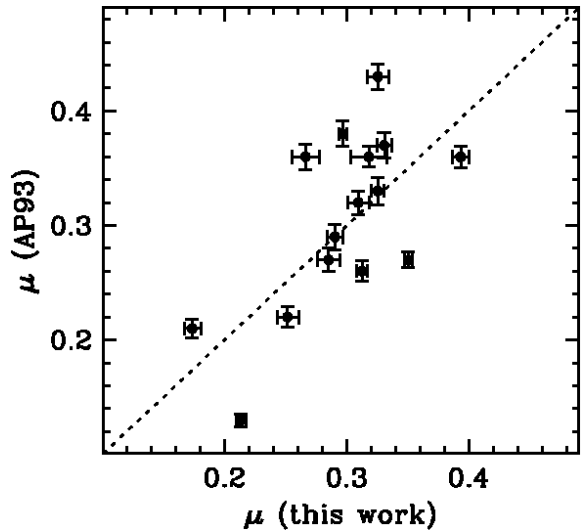}
  \caption{Comparison of proper motion of 15 knots in common with
    \citet{anup93}.
\label{F-companup}}
\end{figure}

\citet{anup93} (AP93) have measured the motion in the plane of the sky
of 20 individual bright knots using two ground based images obtained
on 1984.63 and 1990.10 at Vainu Bappu Observatory (India), with a
similar filter as in our observations.
From visual inspection of their Figure~1, it is possible to
safely identify 15 knots in common with our measurements.  The
comparison of their proper motions is presented in
Figure~\ref{F-companup}. Individual differences between the two
datasets are above the quoted errors, but no systematic effects
appear. Given that our measurements come from the analysis of 19
accurately registered high-quality images, we keep our proper motion
measurements for the following analysis.

\section{RADIAL VELOCITIES}\label{S-rv}

\begin{figure*}[!ht]
\epsscale{2.30}
\plottwo{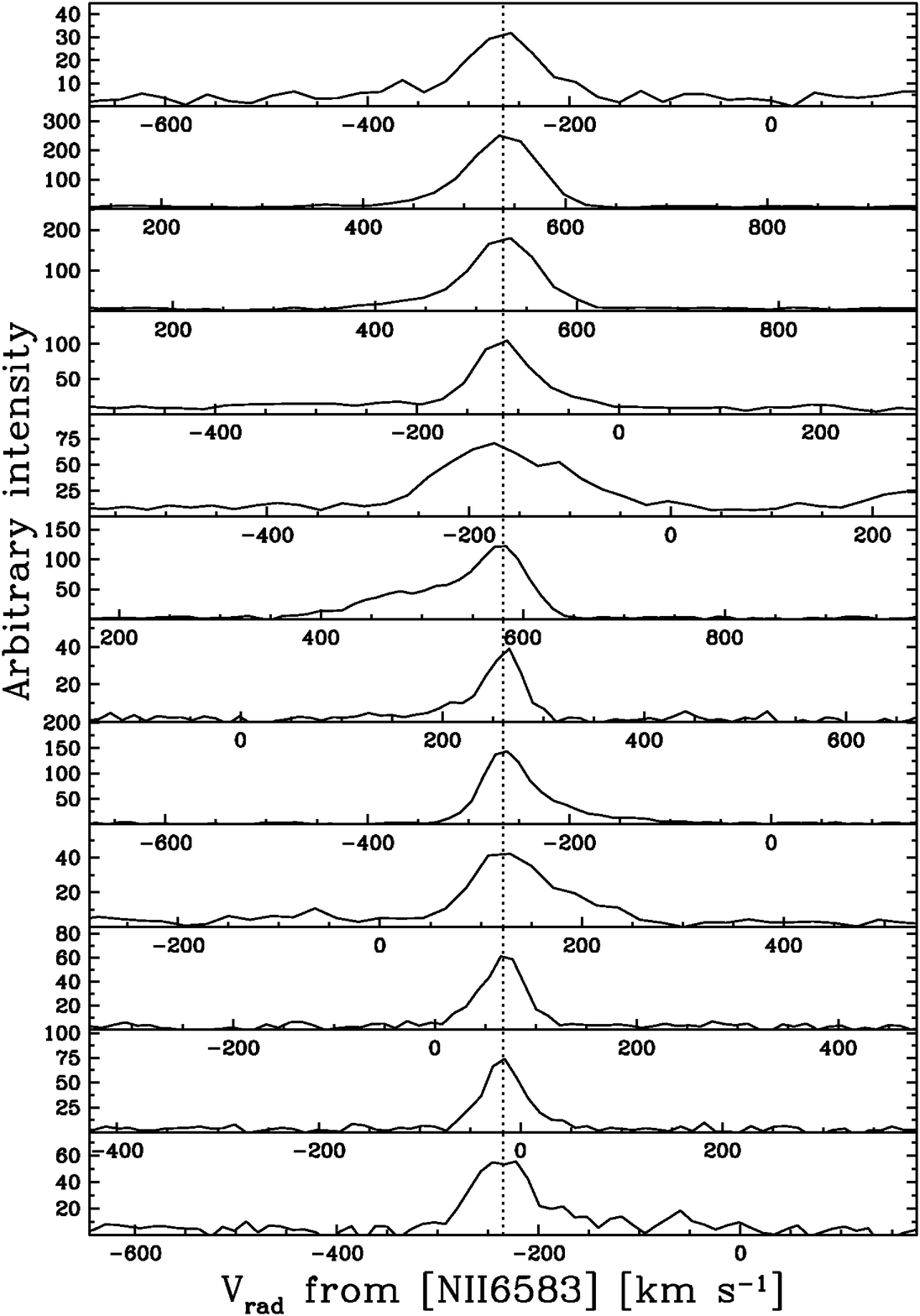}{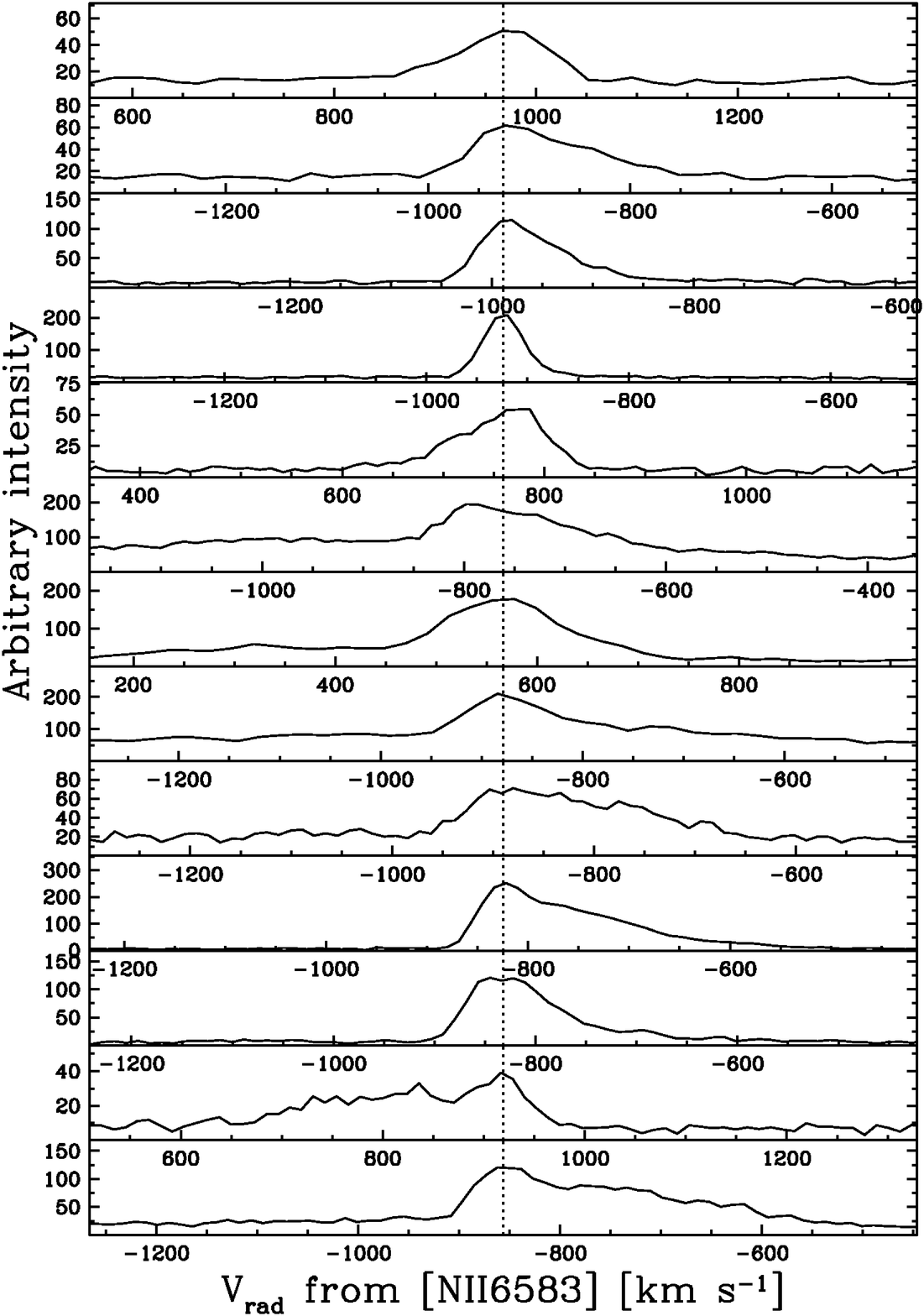}
  \caption{Example of intensity profiles of knots on the edges (left panel) and 
  near the center (right panel) of the remnant.
    Vertical dotted line indicates the measured radial velocity for the
    corresponding knot. \label{F-ip}}
\end{figure*}

Our medium resolution spectra include the H$\alpha$ line and the
[NII] doublet. No other emission lines were detected at the depth of
these observations in the relatively small spectral range covered 
(see Table~\ref{T-specobs}). First, the spatial regions which
clearly correspond to knots identified in the images obtained at the
same epoch were extracted from the long-slit data. In this way, 217
knots
could be identified and measured.  Then the signal in each region was
integrated to create the H$\alpha$ and [NII] line profiles of each
knot. They are generally broad, as also noted by \citet{sh12}, with an
instrument-corrected full width at half maximum up to 200~\kms, as
illustrated in Figure~\ref{F-ip}.
The figure also shows that knots toward the center of
the nebula (right panel) frequently have marked asymmetric shapes, with tails
extending to velocities as large as 300 \kms. Given the strong
projection effects in these central regions, it is not clear if these
tails are intrinsic to the knots, associated with related filaments,
or caused by apparent superposition of more than one knot.

The radial velocities of each knot were then measured by Gaussian
fitting the [NII] 6583~\AA\ line profile. When the line clearly had
more than one peak or possessed non-negligible wings, multi-Gaussian
fitting was applied, but the secondary components (which takes into
account the wings or minor peaks) were not used in the following
kinematical analysis.  Errors on the radial velocity have been
estimated by comparing multiple measurements of the same knots in
different nights and with the two different telescopes used. A typical
error of $8$~\kms~is adopted.  All velocities were corrected to the
Local Standard of Rest (LSR). As the LSR systemic velocity of GK~Per
was estimated by \citet{bode04} to be $45\pm4$~\kms, radial velocities
were also corrected by this value. Table~\ref{T-propm} lists the
radial velocities of knots with respect to the systemic velocity of GK
Per. They range from $-989$ to $+967$~\kms.

\section{KINEMATICAL ANALYSIS}\label{S-kinmod}

\begin{figure}[!ht]
\epsscale{1.0}
\plotone{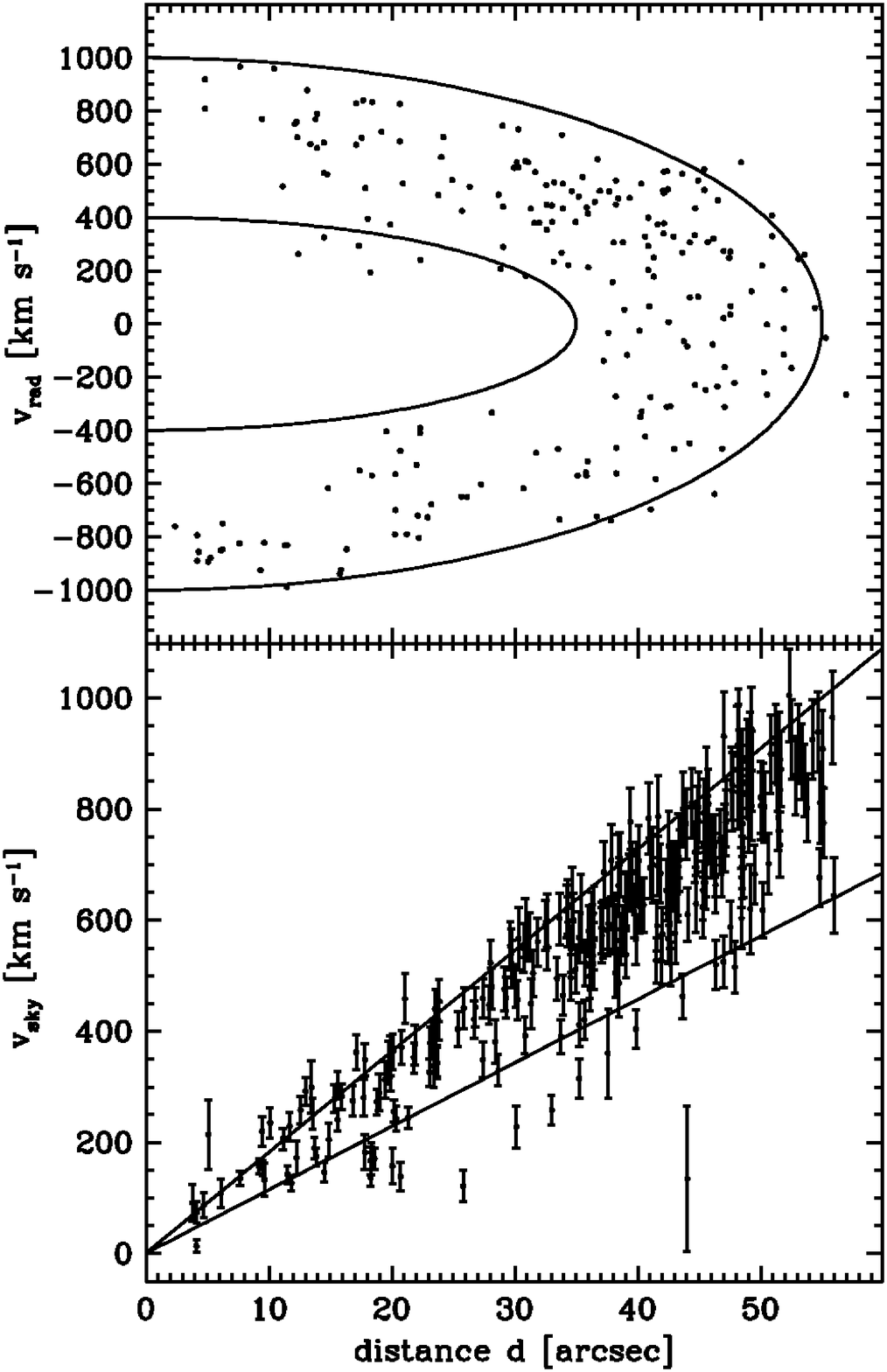}
\caption{Above: Doppler-shift velocities of 217 knots as a function of
  their distance from the central star projected in the plane of the
  sky. Solid lines indicate the position-velocity plot for thin
  spherical shells with the parameters indicated in the text. Below:
  the same, but for the velocity component in the plane of the sky
  from the measurements of the proper motions of 282
  knots.\label{F-pm_vrad}}
\end{figure}

Given its roundish overall morphology, as a first approximation we
consider the nebula of GK~Per to be a quasi-spherical shell in which
knots expand radially from the central star.  This hypothesis is
broadly consistent with the plot of the observed radial velocities
$v_{rad}$ of the knots as a function of their apparent distance from
the central star (Figure~\ref{F-pm_vrad}, upper panel).  The two solid
lines in the figure indicate the expected distance-velocity plot for
spherical shells of radii of 35 and 55 arcsec, and expansion
velocities of 400 and 1000~\kms, respectively. Most of the knots are
comprised within these two limits, suggesting that the knots of the
nebula of GK~Per form a relatively thick shell expanding with a
significant range of velocities.

This conclusion is supported by the behavior of the proper 
motions $\mu$.  The component of the expansion velocity in the plane
of the sky immediately follows from $\mu$ according to equation (in
convenient units)
\begin{equation}
v_{sky}\ [km\ s^{-1}] = 4.74 \cdot \mu\ [''\ yr^{-1}] \cdot D\ [pc],
\end{equation}
where $D$ is the distance to GK~Per.  For a distance of $400\pm30$ pc
(Section \ref{S-kinmod}), $v_{sky}$ of the measured knots varies
between 13 and 1005~\kms, and its variation with 
distance from the center is also consistent with a
spherical thick shell model with parameters as described above
(Figure~\ref{F-pm_vrad}, lower panel). 

\begin{figure}[!t]
\plotone{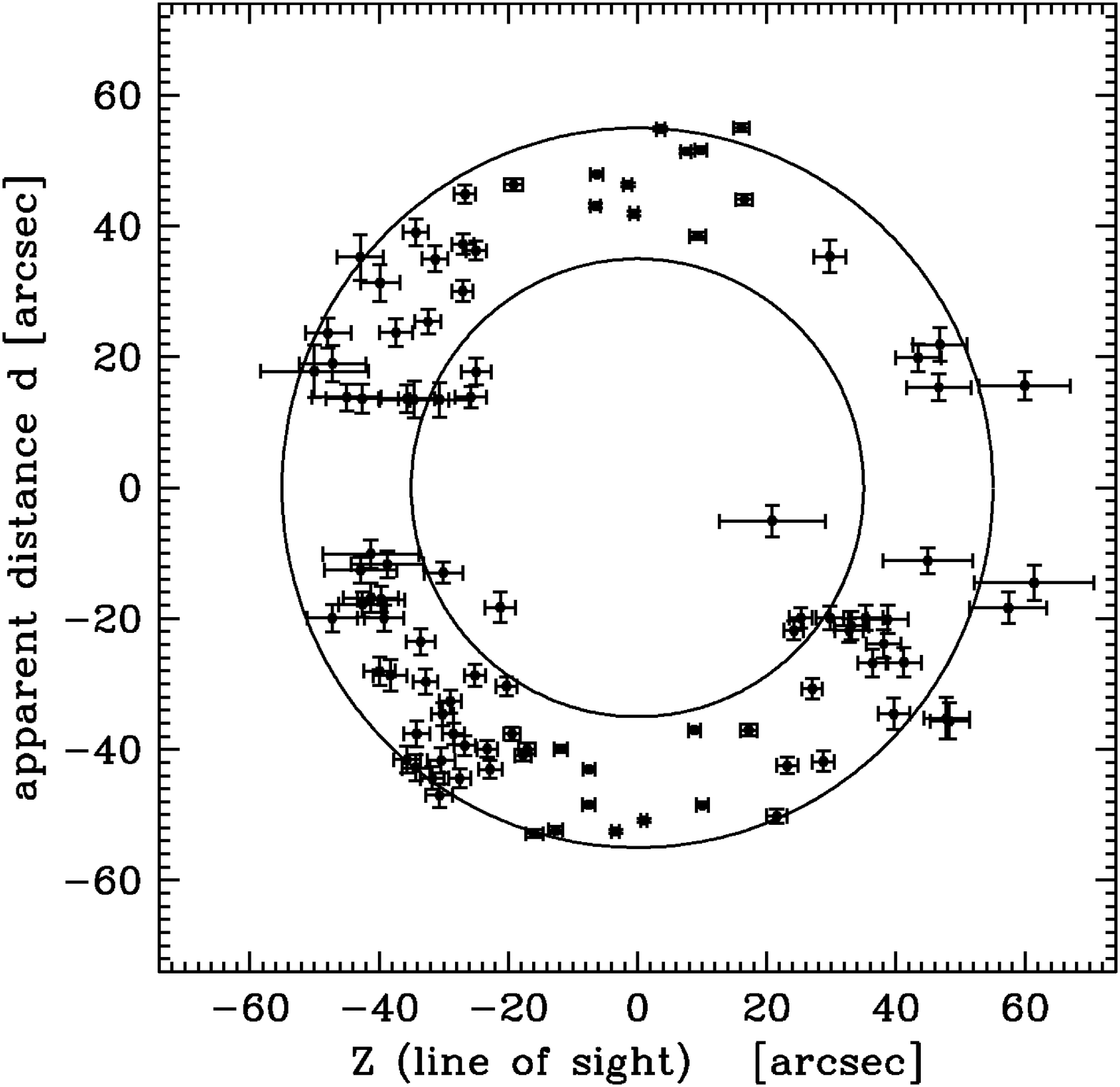}
  \caption{Nebular geometry along the line of sight ($Z$ direction,
    observer on the right side) for the adopted distance to GK~Per of
    400~pc.  Circles indicate inner and outer radii as in
    Figure~\ref{F-pm_vrad}. See text for details.\label{F-foldcut}}
\end{figure}

The combination of the velocity components in the plane of the sky and
along the line of sight provides the velocity vectors of the knots 
(see the right panel of the Figure~\ref{F-geom})
once the distance to GK~Per is fixed.  We could measure both velocity
components for 117 knots:
their deprojected speed is $v_{exp} = \sqrt{v_{sky}^2+v_{rad}^2}$, and
$\theta_{1}=\arctan(v_{sky}/v_{rad})$ is the angle between the line of
sight and the velocity vector (as in Figure~\ref{F-geom}). Assuming a
purely radial expansion, $\theta_{1}$ is also the angle that the line
joining the central star with a knot forms with the line of sight,
$\theta_{2}$ in Figure~\ref{F-geom}.  The deprojected distance from
the center would therefore be $R=d/|\sin\theta_1|$, and its projection
along the line of sight is $Z=R\cos\theta_1$.  We define the sign of
$Z$ so that it is positive when a knot is approaching us
(i.e. blue-shifted) and negative when it is receding (red-shifted). We
also define $d$ to be positive for all knots in the Northern side of
the nebula and negative for those in the Southern half. With these
notations, a representation of the ``depth'' of the nebula along the
line of sight (i.e. its extension along the $Z$ direction) is shown in
Figure~\ref{F-foldcut}. In the figure, the observer is on the right
side, and the plane $Zd$ represents the folding of all planes
perpendicular to the plane of the sky and passing through the central
star and the knots considered  
(right panel of Figure~\ref{F-geom}).  99 knots with a computed error on
$R$ smaller than 10 arcsec are plotted. As stated above, the extension
of the nebula along $Z$ depends on the adopted distance $D$. We assume
that the distance value that better corresponds to an overall
spherical shape of the nebula is our ``expansion-parallax''
determination of the distance to GK~Per. The adopted value, estimated
by visual inspection of the variation of the overall shape of the
nebula as a function of $D$, is 400$\pm$30~pc.  This is broadly
equivalent to assuming that $v_{sky}$ for knots at the outer apparent
edge of the nebula is the same as $v_{rad}$ for knots in the innermost
regions.

\begin{figure}[!t]
\plotone{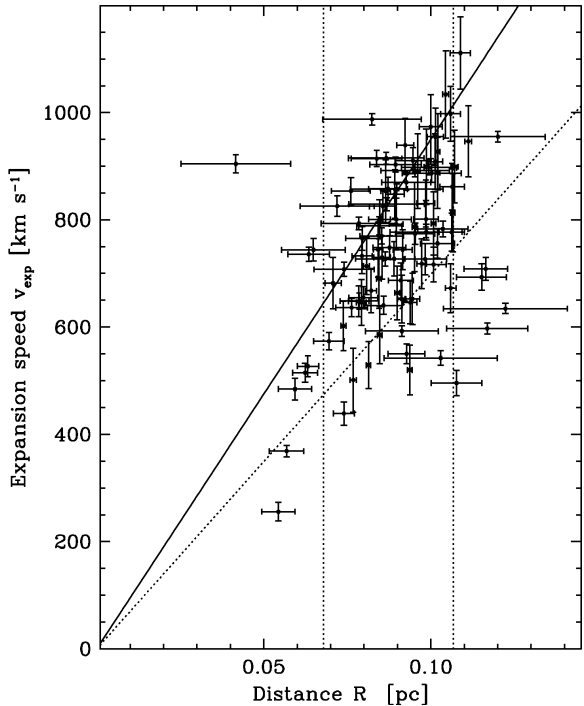}
  \caption{Expansion speed vs. linear distance from the central star
    for $D$=400~pc. The dotted vertical lines indicate the radius
    limits of 35 and 55 arcsec as in the previous graphs. The solid
    line indicates the relation expected for ballistic expansion of
    material ejected in 1901 (102.9~years from our reference
    epoch). The dotted line shows the same relation for an age of 140
    years.
\label{F-hubble}}
\end{figure}

Figure~\ref{F-hubble} shows the relations between deprojected
quantities, namely the speed $v_{exp}$ and the distances $R$ from the
central star. As in previous graphs, it confirms that the nova 
remnant is thick (nearly a half of its outer radius) and that the 
expansion speed of knots covers the significant range from 250 to 
1100~\kms, with the majority of them having a speed between 600 and 
1000~\kms.  In the figure, we also indicate the expected
relationship between velocity and distance for a ballistic ejection
(Hubble-like flow) with an age $t_{nova}$=102.9~years, which is the
time lapse between the nova maximum of 1901 Feb 22 and our reference
epoch of 2004 Jan 29. Points above the solid line have a kinematical age
smaller than $t_{nova}$, and those below the line have a larger
kinematical age, which indicates that they suffered some deceleration
in the course of the expansion. For comparison, in the figure we show
as a dotted line also the relation expected for a ballistic expansion
with an age of 140~years. The knots age is further investigated in the
next section.

\subsection{Kinematical ages}\label{S-age}

A direct estimate of the kinematical age $t$ of each knot can be obtained
from the proper motion measurements, according to 
\begin{equation}
t\ [yr]= \frac{ d\ [''] } {\mu\ [''\ yr^{-1}]},
\end{equation}

where $d$ is the distance of the knot from the central star on the
plane of the sky that we measured at our reference epoch (see
Figure~\ref{F-geom}).

\begin{figure}[!t]
\plotone{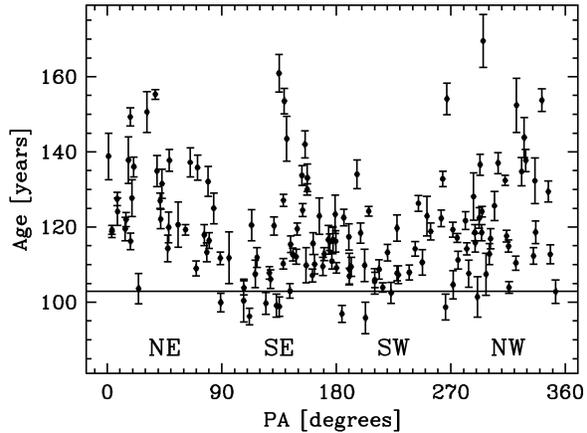}
  \caption{Kinematical ages as a function of the knots position
    angle, for knots with $d\ge35''$. The horizontal line is the real age of the nova, 102.9 years. 
\label{F-agepa}}
\end{figure}

In Figure~\ref{F-agepa} we present the kinematical age computed in
this way as a function of the knots position angle.  In the figure,
only the 151 knots with an estimated error on the age smaller than 8
years, and with an apparent distance from the centre
$d\ge$~35~arcsec (in order to focus on the directions close to the
plane of the sky and minimize projection effects) are included.  Ages
vary between 96 and 170 years, the mean value, weighted by errors,
being $118\pm12$ years. The horizontal line indicates the nova age
$t_{nova}$, and shows that in general knots have suffered only a 
modest deceleration since their ejection. This conclusion is
consistent with the comparison between the knots speeds of up to
1100~\kms\ determined above (Figure~\ref{F-hubble}), and an initial
expansion of the outflow of $\sim$1300~\kms\ \citep{p59}.  Indeed,
in the simplistic hypothesis of constant deceleration throughout the
knots lifetime \citep{du87}, their observed mean age of 118 years
implies that the present-day average speed is 77\%\ of the original
ejection velocity. The time $t_{\frac{1}{2}}$ for individual knots to
decrease in speed by a factor of two would be 220 years.  Even for
knots with an apparent age of 140 years (the upper limit for the vast
majority of points in Figures~\ref{F-hubble} and \ref{F-agepa}),
velocity would have decreased by only 42\%\ in the 102.9 years since
the nova explosion, and $t_{\frac{1}{2}}$ would be 120 years.  Data
therefore exclude the short timescale $t_{\frac{1}{2}}$=58~years
estimated by \citet{du87}.

The calculations of \citet{du87} were based on the measurement of the
radius of the SW side of the nebula from images taken between 1917 and
1984. These measurements were extended by \citet{seaq89}, who analyzed
them using a more physical model which considered the expansion of an
isothermal or adiabatic shock created by the interaction of the
expanding outflow with a tenuous circumstellar medium. Such an analysis
was further extended by \citet{anup05} with images taken till 2003.
We refrain from further extending the method, as we found that the
measurement of the size of the nebula is highly subjective, owing to
the difficulty in defining the nebular edge due to the lack of an abrupt
fall of surface brightness, clumpiness (which is enhanced by the
better resolution of the recent images), and asymmetries. Indeed, 
our estimates of the size differ from those of \citet{seaq89} and
\citet{anup05} by up to several arcseconds, which prevents any
detailed discussion or refinement of the results obtained by these
authors.

\subsection{Deviations from spherical symmetry}

The proper motion measurements also allow us to measure any deviation
from circular symmetry of the expansion pattern in the directions
close to the plane of the sky. Anisotropies along the line of sight
depends on the adopted distance of the system, and indeed in the
previous section we used the argument inversely, having minimized any
anisotropy in order to estimate a distance to the system.

Considering the data in Figure~\ref{F-agepa}, the mean weighted age in
the different quadrants is: $125\pm14$ yr (NE, 34 knots), $115\pm11$
yr (SE, 45 knots), $113\pm10$ yr (SW, 34 knots), and $121\pm11$ yr
(NW, 38 knots).
We conclude that no strong evidence exists for deviations from the
circular symmetry in the expansion pattern. If any, knots in the
Northern side of the nebula seem to have suffered a stronger
deceleration throughout the remnant lifetime than those in the
Southern part. Therefore, our calculations do not support the
prediction of \citet{bode04} that knots in SW quarter should have been
slowed down more than the rest. Smaller proper motions in the NW
quadrant are also mentioned in \citet{sh12}.

\begin{figure}[!t]
\plotone{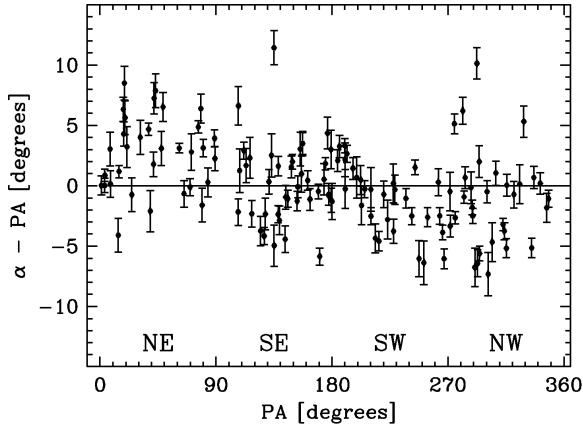}
  \caption{Deviations of the proper motion velocity vector $\alpha$
    from the position angle PA of knots.
\label{F-alphpa}}
\end{figure}

In addition, our initial hypothesis of purely radial motions can be
tested by comparing the direction of the velocity vectors in the plane
of the sky, $\alpha$, with the position angle PA of knots in the plane
of the sky, i.e. their radial direction from the central star. As in
Figure~\ref{F-agepa}, only knots with $d\ge$35~arcsec are considered
in order to limit projection effects.  Figure~\ref{F-alphpa} shows
that velocity vectors are generally aligned along the radial direction
but a clear symmetric pattern of non-radial velocities is also
observed. The maximum deviations are observed in the NE quadrant,
where knots trajectories seems systematically bended toward the East
with respect to the radial direction, and the Western part of the
nebula, where knots velocity vectors are preferentially bended toward
the South.

\section{BRIGHTNESS EVOLUTION}\label{S-br}
 
\begin{figure}[!t]
\plotone{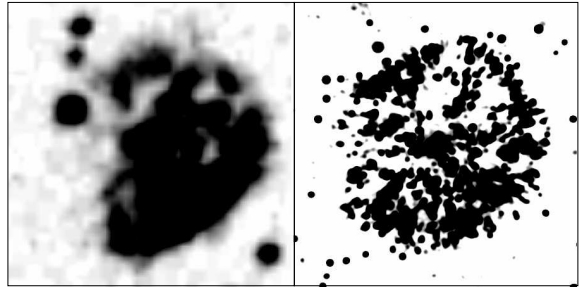}
  \caption{Comparison of the 1953 photographic image of GK~Per 
    with a smoothed version of our 2011 CCD image. The former is displayed 
    with a $\times$2 zoom with respect to the latter for easier comparison.
\label{F-1953vs2011}}
\end{figure}

The GK~Per optical nebula suffered clear morphological and brightness
changes in the last 50 years.  This is illustrated in
Figure~\ref{F-1953vs2011}, where a $\times$2 zoom of the 1953 image is
compared to a smoothed version of our 2011 image.  The main changes
are a relative brightening of the whole Eastern side of the nebula
compared to the Western side, and in particular the development of a
more prominent ``bar'' in the NE edge (see also \citealt{anup05}), and
a progressive circularization of the SW edge of the nebula, i.e. of
the region of the strong interaction observed at radio and X-ray
wavelengths. In general, the outflow looks more circular and uniform
than in the past.

\begin{figure}[!t]
 \plotone{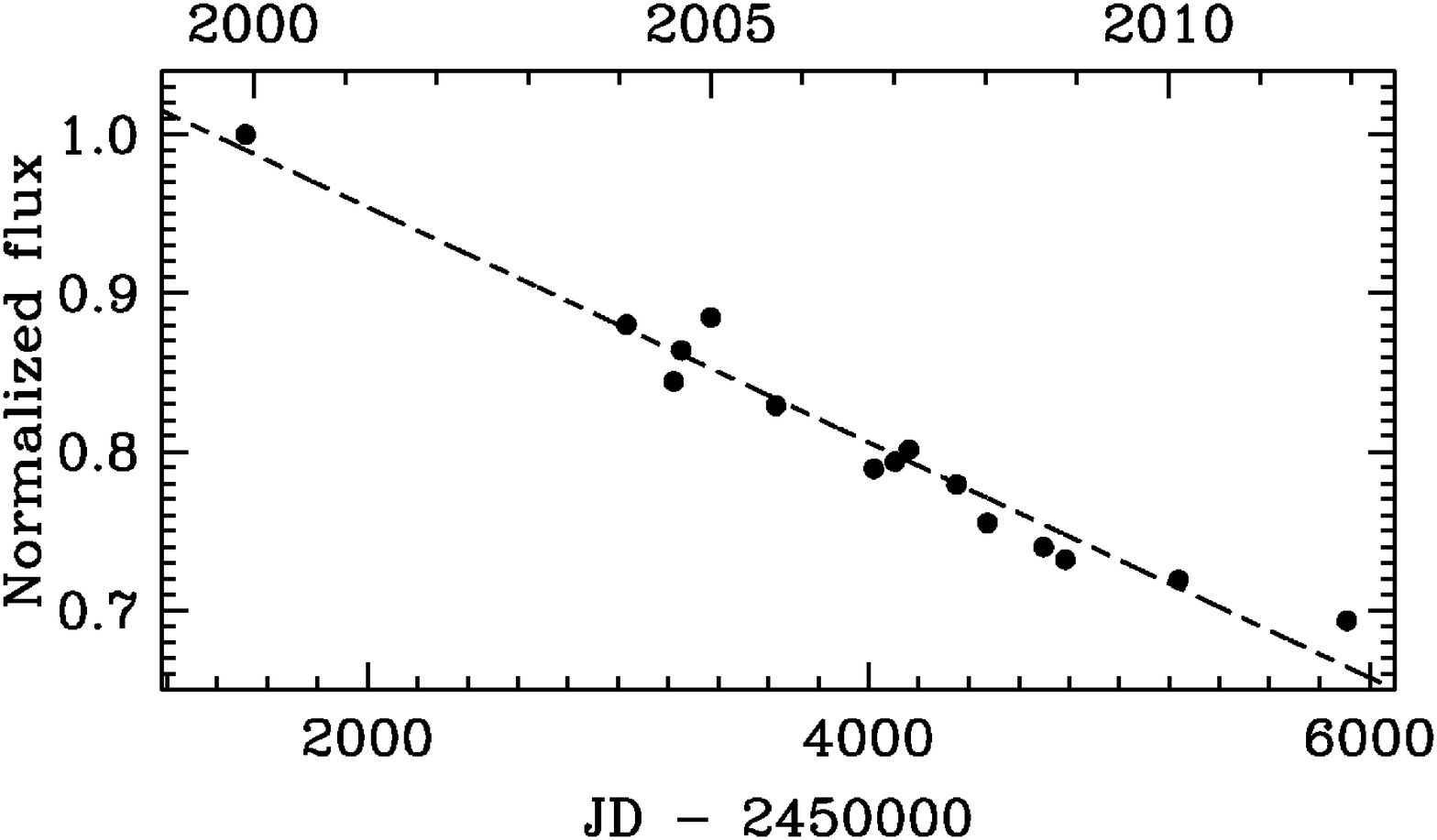}
  \caption{Total $H\alpha$+[NII] flux of the remnant from 1999 to 2011.\label{F-totflux}}
\end{figure}

As most of our images were taken with the same instrument, detector,
and filter, the brightness variation of the nebula in the last decade
can be quantified.  The 15 INT+WFC flux-matched frames marked with an
asterisk in Table~\ref{T-imaobs} were used.  To measure the brightness
of the nebula, all stars inside or nearby the remnant were masked.
We first measured the brightness variation of the nebula as a whole.
In Figure~\ref{F-totflux} we present the evolution of the total flux
of the remnant normalized to the value that it had in 1999.  It shows
that the total $H\alpha$+[NII] flux of the nebula has been linearly
decreasing, for a total of 31\%\ over the 12.04 years considered,
i.e. at a rate of 2.6\%\ per year, which is similar to what measured
in the radio by \citet{anup05} and attributed to adiabatic expansion
of the remnant into the surrounding medium.

\begin{figure}[!t]
\plotone{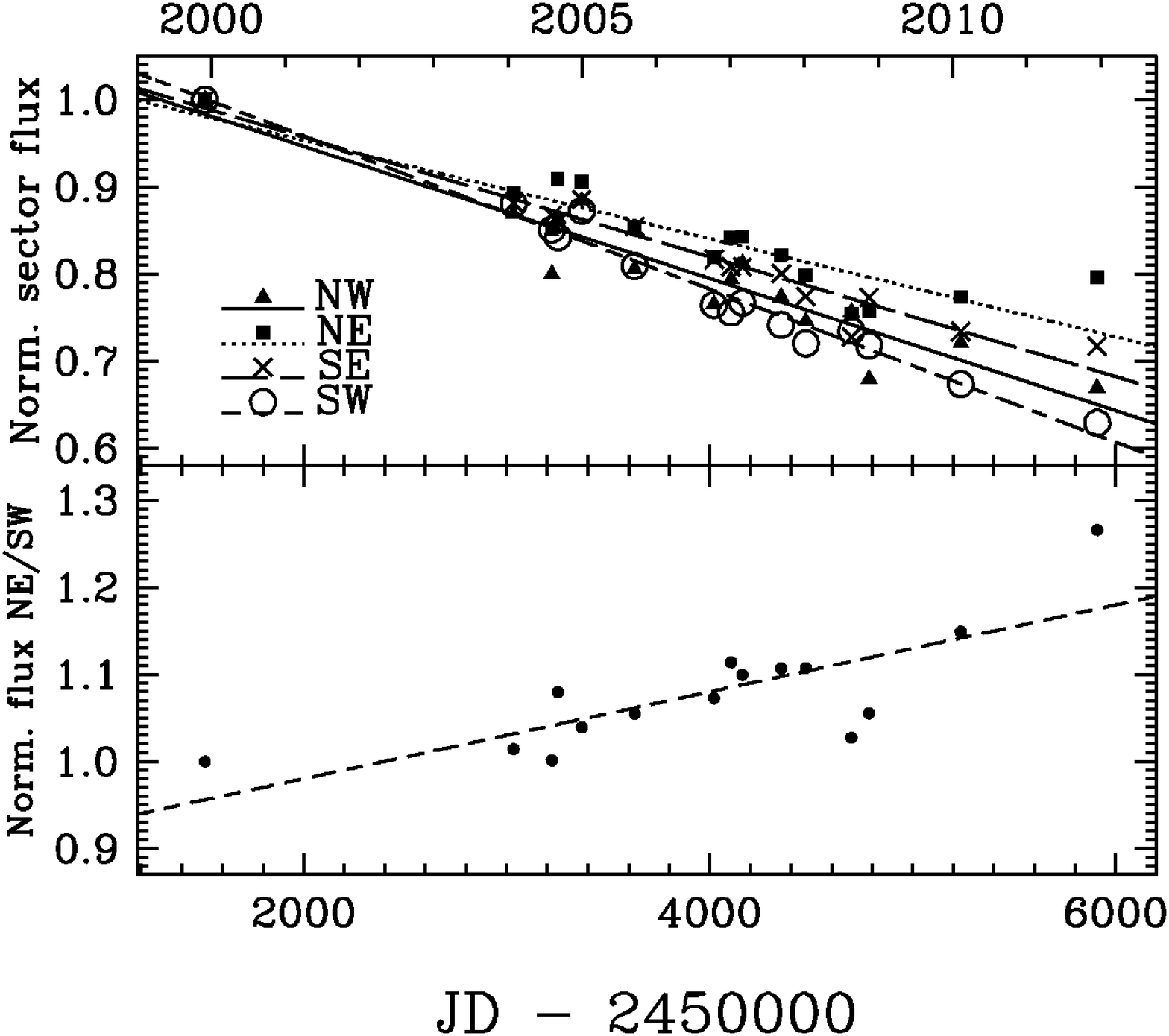}
  \caption{Top panel: flux variations for different quadrants. 
    Lower panel: flux ratio variations of NE and SW
    quadrants.\label{F-secflux}}
\end{figure}

Second, the flux variation for each of the four quadrants (NE, NW, SE,
SW) was computed. As can be seen from Figure~\ref{F-secflux}, the NE quadrant shows the
shallowest decline, with even some possible re-brightening in the last
3 years. The quadrant fading at the largest rate is the SW one.

\begin{figure}[!t]
\plotone{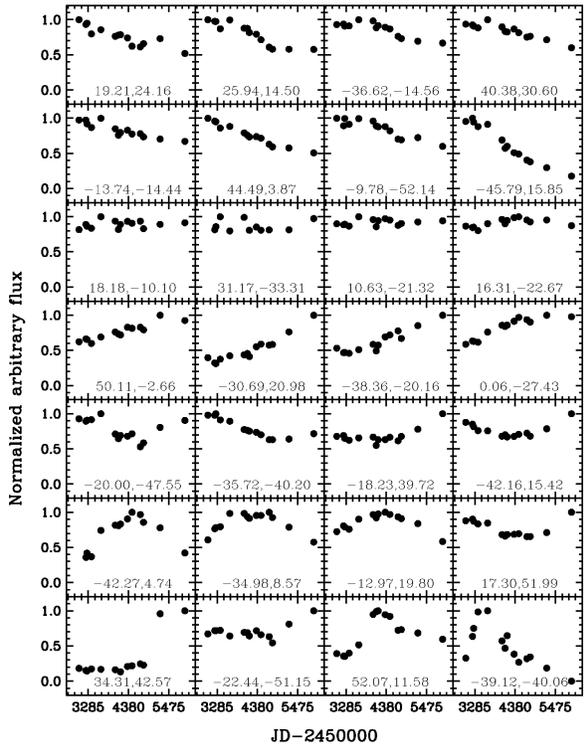}
  \caption{Illustrative example of the brightness changes of
    individual knots. The knots $d_{x}$ and $d_{y}$ coordinates as in
    Table~\ref{T-propm} are indicated.\label{F-bright}}
\end{figure}

Third, we have also considered the flux variations for a number of
individual knots. Their fluxes were measured using circular apertures
matched to the knots size, using an external annulus for background
subtraction.
The 1999 image could not be used for the measurement of individual
knots because of its poorer seeing.  After excluding knots in crowded
areas, we could measure the flux variation of 85 knots. Illustrative
examples are presented in Figure~\ref{F-bright}. In the figure, the
flux is normalized to the maximum flux of each knot in the period
considered.  
Out of 85 knots, 50\%\ show a similar fading as the nebula as a whole
(first two rows in the figure).  Some 15\%\ have a constant flux
(third row), while some 20\%\ of the knots show instead a monotonic
brightening (fourth row in Figure~\ref{F-bright}). Finally, 15\%\ of
the knots are rapidly fading or brightening (or both) during the
2004-2011 period (last three rows). The most extreme cases show a
sudden brightening of a factor of 5 in one year (bottom-left panel),
or a similarly fast brightening and then a dramatic fading on a
timescale of less than three years (bottom-right panel).  The
brightness variations of individual knots is also discussed by
\citet{sh12}.

\section{DISCUSSION}\label{S-disc}

\subsection{Distance}

Our kinematical determination of a distance of 400$\pm$30~pc for
GK~Per (Section~\ref{S-kinmod}) compares well with the value derived
using the absolute peak luminosity of novae. We have considered four
different relationships between $M_V$ at peak luminosity and the rate of the 
nova decline (\citealt{coh85,cap89,del95,dd00}), depending on $t_2$ 
(the time in days the nova takes to fade 2 mag from the peak brightness). Assuming the
following for GK~Per: $t_2=7$ d, a peak visual magnitude of $m_v$=0.2 mag, and
a reddening of $E(B-V)=0.3$ (\citealt{wu89}), 
we have computed the distance to GK~Per using
the different relationships. The average distance is 440$\pm$70~pc, in good
agreement with our kinematic estimate.

Our distance also broadly agrees with (and should improve) the
estimates from other authors using the expansion parallax of the
remnant.  The most widely used distance is 470 pc by \citet{mcl60}.
\citet{du81} found the distance to be 525 pc. A more recent
determination by \citet{sl95} states $455\pm30$ pc.

Finally, a rather short distance to GK~Per, 337 pc, was determined in
\citet{war87} using the Bailey's method (\citealt{bai81}). This method
considers the $K$-band magnitude of the secondary star, which for
cataclysmic variables is not easily separated from the disc
contribution: therefore the calculated distance should be considered
as a lower limit.  The largest distance determination is the one by
\citet{sh83}, 726 pc, using again the $K$ mag and their estimated
spectral type for the secondary star (K2 IV).

\subsection{Dynamics of the ejecta}

Our kinematical study, combining velocity measurements both in the
plane of the sky and along the line of sights, has highlighted several
basic properties of the dynamical evolution of the GK~Per nova
remnant. The outflow consists of a thick shell composed of about a
thousand knots and filaments. Knots velocities range from 250 to
1100~\kms, with the majority of them having a velocity between 600 and
1000~\kms.  The kinematical ages of knots from the proper motion
measurements indicate that most of them have been only modestly slowed
down in the 102.9 years from the ejection to the present, with no
significant deceleration during the last 10-20 years. Clear 
evidence for systematic deviations of the expansion rate with
direction is either not found.

On the other hand, some long-term evolution of the overall
characteristics of the nebula is highlighted. The most notable changes
are a progressive circularization of the SW nebular edge, and a
relative brightening of the whole Eastern side compared to the Western
part, with the development of a ``bar'' in the NE edge. 

In few words, the nebula seems to expand nearly ballistically, and is
progressively becoming rounder and more uniform in terms of surface
brightness.  These results are somewhat surprising considering the
previous interpretation of the evolution of the GK~Per remnant. In
fact, the asymmetry observed at radio and X-ray wavelengths
\citep{seaq89,bal05} seems to indicate that along the SW direction
strong interaction of the remnant with relatively dense circumstellar
material has occurred. This is however not reflected in enhanced
deceleration of the optical knots along this direction.

The interpretation of the nebular dynamics should take into account
these new results.  A basic role in the discussion is played by the
original circumbinary density distribution into which the nova ejecta
expanded. Its most visible expression is the large bipolar nebula
first detected from IRAS data by \citet{bode87}.  The strongest
argument for the association between this bipolar structure and the
nova is geometrical, that is, the emission is symmetrically placed
around GK~Per. The inherent problems of the classification of the
bipolar nebulosity as an old planetary nebula from the primary star of
the system were presented by \citet{tw95}.  An alternative solution
has been proposed by \citet{doug96}: in it the secondary star could be
responsible for the formation of the extended nebula as it evolved off
the main sequence triggering a major transfer of material into the
primary star causing a born-again asymptotic giant branch event.  As
revealed by deep optical imaging the extended emission presents a
typical bipolar morphology with the polar axis oriented in the NW-SE
direction. The nebula shows an asymmetry between the NE and SW edges
that \cite{bode04} interpreted as the motion of the nebula to the SW
assuming high relative velocity with respect to the ISM. Within this
picture the non-thermal radio \citep{anup93} and the X-rays
\citep{bal99} emission in the SW direction is interpreted by
\cite{bode04} to be a consequence of the nova ejecta ploughing the
denser medium of the SW side of the bipolar nebula.

We can test this idea in the framework of the existing hydrodynamical
models of nebular interaction with the ISM.  The central binary is
moving with a space velocity of $45 \pm 4$ km~s$^{-1}$ \citep{bode04},
or 4.6$\times10^{-5}$ pc~yr$^{-1}$.  For simplification, we consider
the central source as a single star (as successfully done for Mira,
see \citealt{mar07}).  Moreover, given that the direction of the
movement is almost perpendicular to the axis of the bipolar lobes we
can also assume spherical symmetry (the formation of a bipolar
structure is complex and beyond the scope of the present work). With
this simplifications at hand we can consider that the interaction
process can be similar, at least regarding the discussion that
follows, to that of the wind of an AGB star with the ISM (see e.g.
\citealt{villa03, villa12}). In particular, a simulation with a
relative speed of $50$~\kms\ can be found in \cite{villa12}. These
simulations show that once a bow-shock structure is formed, the
subsequent stellar wind expands unperturbed inside the asymmetric
outer cavity created by the interaction.  In these conditions, never
the stellar material ejected inside this cavity can feel the outer
interaction if it is inside the bow-shock structure. The same idea
could be applied to GK~Per: if both the outer bipolar nebula and the
nova ejecta are the result of the evolution of the central binary, and
the nova shell is expanding inside the bipolar nebula, then the ejecta
cannot see the interaction with the ISM, and should expand
unperturbed.  This interpretation is consistent with the lack of
evidence for deviations of the expansion rate with direction found in
this work. In particular, the SW side, where most of the interaction
with the circumstellar medium was assumed to occurred in the past (see
radio and X-ray data) does not show any sign of larger deceleration
than the rest of the nebula.  The asymmetry observed in the radio and
X-rays, however, would remain unexplained in this scenario.

An alternative possibility is the existence of a mechanism that allows
interaction with the ISM to permeate from the outer structure to the
nova shell.  An interaction with magnetized ISM could provide this
mechanism \citep{sok97}. In fact, the nature of the radio and X-ray
emission could be explained as dense gas clouds moving supersonically
through a magnetized low-density medium.  \cite{jon94} show how the
synchrotron emission from relativistic electrons increases rapidly
when the cloud begins to fragment under a Rayleigh-Taylor
instability. They also show how the interaction of this wind with
dense condensations can lead to strong magnetic fields. If
relativistic particles are formed as the fast wind is shocked, then
the enhanced magnetic field will result in non-thermal radio emission.

At this point we can turn our attention back to the outburst mechanism
that leads to the formation of the nova ejecta.  Its thick shell
geometry can be the result of an ejection with a range of initial
velocities, or the dynamical evolution of the interaction of different
components.  We favour the latter hypothesis, which is supported by the
observations of nova shells soon after outburst. It has been argued in
the literature that two distinct outflow components of different
origin are present at early times: an earlier, slow one from the
circumbinary gas, and a later, fast one result of the outburst (see
e.g. \citealt{cass04,Williams10}).  In the case of Nova Cyg 1992,
\citet{cass04} show that the main shell ejected at the moment of the
outburst contains the bulk of the ejected matter and is located in the
outermost region of the ejecta region, while a less massive impulsive
event forms a faster expanding shell that eventually will catch up
with the main shell generating a region of shock interaction where the
two shells meet. Eventually, the process gives rise to the formation
of a new shell formed by the merger of the two. A similar scenario in
terms of discrete shell ejection events is proposed by \citet{wil08}
and \citet{Williams10} to explain the absorption line spectra of the
early ejecta of 15 nova shells.

In GK~Per we are probably witnessing the late evolution of such of a
process, and in this respect it is natural that any asymmetry
originally present in the dense ejecta (see e.g. the 1917 images in
\citealt{bode04}) disappears with time as the combined shell expands
outward. This explains the progressive circularization and
homogenization of the optical outflow. The original asymmetry can
also be understood in terms of the binary interactions in the system
(e.g. \citealt{syt09}).

\acknowledgments

Based on observations made with the Isaac Newton Telescope, operated
by the Isaac Newton Group, and with the Nordic Optical Telescope,
operated jointly by Denmark, Finland, Iceland, Norway, and Sweden, on
the island of La Palma in the Spanish Observatorio del Roque de los
Muchachos of the Instituto de Astrof\'\i sica de Canarias. INT data
were in part reduced by the IAC support astronomer staff.  This paper
also makes use of data obtained from the Isaac Newton Group Archive
which is maintained as part of the CASU Astronomical Data Centre at
the Institute of Astronomy, Cambridge.  This work makes use of EURO-VO
software, Aladin and TOPCAT. The EURO-VO has been funded by the
European Commission through contracts RI031675 (DCA) and 011892
(VO-TECH) under the 6th Framework Programme and contracts 212104
(AIDA) and 261541 (VO-ICE) under the 7th Framework Programme.
Data presented here is partially based on
photographic data of the National Geographic Society -- Palomar
Observatory Sky Survey (NGS-POSS) obtained using the Oschin Telescope
on Palomar Mountain.  The NGS-POSS was funded by a grant from the
National Geographic Society to the California Institute of Technology.
The plates were processed into the present compressed digital form
with their permission.  The Digitized Sky Survey was produced at the
Space Telescope Science Institute under US Government grant NAG
W-2166. 
The compressed files of the "Palomar Observatory - Space Telescope
Science Institute Digital Sky Survey" of the northern sky, based on
scans of the Second Palomar Sky Survey are copyright (c) 1993-2000 by
the California Institute of Technology and are distributed herein by
agreement.  This research has made use of the USNOFS Image and
Catalogue Archive operated by the United States Naval Observatory,
Flagstaff Station (http://www.nofs.navy.mil/data/fchpix/).

We thank Gavin Ramsay for obtaining the 20091208 INT data. 
We also thank Rafael Barrena Delgado for kindly providing us
pre-reduced and combined images for the 20111213 observations. 
The work of RLMC and MSG has been supported by
the Spanish Ministry of Science and Innovation (MICINN) under the
grant AYA2007-66804. PRG is supported by a Ram\'on y Cajal fellowship 
(RYC--2010--05762). EV acknowledges the support provided by the Spanish 
MICINN under grant AYA2010-20630 and to the Marie Curie FP7-People-RG268111.
This work was partially supported by the Spanish MICINN 
under the CONSOLIDER--INGENIO 2010 Program, through the grants
"Molecular Astrophysics: The Herschel and ALMA Era ASTROMOL" 
(CSD2009--00038) and "First Science with the GTC" (CSD2006--00070). 
TL, KV, and IK acknowledge the support of the Estonian Ministry for
Education and Science.





\begin{thebibliography}{}


\bibitem[Anupama \& Prabhu(1993)]{anup93} Anupama, G.~C., \& Prabhu, T.~P. 1993, \mnras, 263, 335

\bibitem[Anupama \& Kantharia(2005)]{anup05} Anupama, G.~C., \& Kantharia, N.~G. 2005, \aap, 435, 167

\bibitem[Bailey(1981)]{bai81} Bailey, J. 1981, \mnras, 197, 31

\bibitem[Balman \& \"Ogelman(1999)]{bal99} Balman, {\c S}., \"Ogelman, H.~B. 1999, \apj, 518, L111

\bibitem[Balman(2005)]{bal05} Balman, {\c S}. 2005, \apj, 627, 933

\bibitem[Barnard(1916)]{bar16} Barnard, Frederick A.~P. 1916, Harvard College Observatory Bulletin, 621, 1

\bibitem[Bode et al.(1987)]{bode87} Bode, M.~F., Seaquist, E.~R., Frail, D.~A., at al. 1987, \nat, 329, 519

\bibitem[Bode et al.(2004)]{bode04} Bode, M.~F., O'Brien, T.~J., \& Simpson, M. 2004, \apj, 600, L63


\bibitem[Capaccioli et al.(1989)]{cap89} Capaccioli, M., della Valle, M., Rosino, L., \& D'Onofrio, M. 1989, \aj, 97, 1622 

\bibitem[Cassatella et al.(2004)]{cass04} Cassatella, A., Lamers, H.~J.~G.~L.~M., Rossi, C., 
Altamore, A., \& Gonz{\'a}lez-Riestra, R. 2004, \aap, 420, 571

\bibitem[Cohen(1985)]{coh85} Cohen, J. G. 1985, \apj, 292, 90 


\bibitem[della Valle \& Livio(1995)]{del95} della Valle, M., \& Livio, M. 1995, \apj, 452, 704 

\bibitem[Dougherty et al.(1996)]{doug96} Dougherty, S.~M., Waters, L.~B.~F.~M., Bode, M.~F.,
et al. 1996, \aap, 306, 547

\bibitem[Downes \& Duerbeck(2000)]{dd00} Downes, R.~A., \& Duerbeck, H.~W. 2000, \aj, 120, 2007 

\bibitem[Duerbeck(1981)]{du81} Duerbeck, H.~W. 1981 in \pasp, vol. 93, 165

\bibitem[Duerbeck(1987)]{du87} Duerbeck, H.~W. 1987,  Ap\&SS, 131, 461



\bibitem[Jones et al.(1994)]{jon94} Jones, T.~W., Kang, H., \& Tregillis, I.~L. 1994, \apj, 432, 194

\bibitem[Martin et al.(2007))]{mar07} Martin, D.~C., Seibert, M., Neill, J.~D., et al., 2007, \nat, 448, 780

\bibitem[McLaughlin(1960)]{mcl60} McLaughlin, D.~B. 1960, Stellar atmospheres, Vol. 6, ed. J. L. Greenstein 
(Chicago: University of Chicago Press), p. 585.


\bibitem[Pottasch(1959)]{p59} Pottasch, S 1959, Ann. d'Astrophys., 22, 297

\bibitem[Ritchey(1902)]{r02} Ritchey, G.~W. 1902, \apj, 15, 129 

\bibitem[Seaquist et al.(1989)]{seaq89} Seaquist, E.~R., Bode, M.~F., Frail, D.~A., et al. 1989, \apj, 344, 805S

\bibitem[Shara et al.(2012)]{sh12} Shara, M.~M., Zurek, D., De Marco O., et al., 2012, \aj, 143, 143

\bibitem[Sherrington \& Jameson(1983)]{sh83} Sherrington, M.~R., \& Jameson, R.~F. 1983, \mnras,, 205, 265

\bibitem[Slavin et al.(1995)]{sl95} Slavin, A.~J., O'Brien, T.~J., \& Dunlop, J.~S. 1995, \mnras, 276, 353

\bibitem[Soker \& Dgani(1997)]{sok97} Soker, N., \& Dgani, R. 1997, \apj, 484, 277

\bibitem[Sytov et al.(2009)]{syt09} Sytov, A.~Y., Bisikalo, D.~V., Kaigorodov, P.~V., \& Boyarchuk, A.~A. 
2009, Astronomy Reports, 53, 223

\bibitem[Tweedy(1995)]{tw95} Tweedy, R.~W. 1995, \apj, 438, 917

\bibitem[Villaver et al.(2003)]{villa03} Villaver, E., Garc{\'{\i}}a-Segura, G., \& Manchado, A. 2003, \apjl, 585, L49

\bibitem[Villaver et al.(2012)]{villa12} Villaver, E., Manchado, A., \& Garc{\'{\i}}a-Segura, G. 2012, \apj, 748, 94

\bibitem[Warner(1987)]{war87} Warner, B. 1987, \mnras, 227, 23

\bibitem[Williams(1901)]{w01} Williams, A.~S. 1901, \mnras, 61, 337

\bibitem[Williams et al.(2008)]{wil08} Williams, R., Mason, E., Della Valle, M., \& Ederoclite, A. 2008, \apj, 685, 451

\bibitem[Williams \& Mason(2010)]{Williams10} Williams, R., \& Mason, E. 2010, \apss, 327, 207

\bibitem[Wu et al.(1989)]{wu89} Wu, C.-C., Holm, A.~V., Panek, R.~J., et al. 1989, \apj, 339, 443 

\end{thebibliography}
\end{document}